\documentclass[aps,jcp,superscriptaddress,amsmath,amssymb]{revtex4-1}

\usepackage{comment}
\usepackage{tikz}
\usetikzlibrary{trees}
\usetikzlibrary{decorations.pathmorphing}
\usetikzlibrary{decorations.markings}
\tikzset{
    photon/.style={decorate, decoration={snake,segment length=1.5mm}, draw=black},
    coulomb/.style={dotted},
    electron/.style={draw=black, postaction={decorate},
        decoration={markings,mark=at position .55 with {\arrow[draw=black]{>}}}}, 
    gluon/.style={decorate, draw=magenta,
        decoration={coil,amplitude=4pt, segment length=5pt}},
    boundelectron/.style={thick, double},
    transverse/.style={dashed}
}

\usepackage{graphicx}% Include figure files
\usepackage{dcolumn}% Align table columns on decimal point
\usepackage{bm}% bold math
\usepackage{rotating}

\newcolumntype{.}{D{.}{.}{8}}

\usepackage[utf8]{inputenc}
\usepackage{physics}
\usepackage{amsmath, amssymb}
\usepackage{tabularx,booktabs,array,dcolumn}
\usepackage{setspace}
\usepackage{vmargin}
\usepackage{xcolor}
\usepackage{graphicx,psfrag,subfigure}

\usepackage{float}
\usepackage[all]{xy}

\usepackage{bm}
\usepackage{mathtools}
\usepackage{physics}
\usepackage[english]{babel}

\usepackage[unicode]{hyperref}
\usepackage{soul}
\newcommand{\bos}[1]{\boldsymbol{#1}}

\newcommand{\cm}{\text{cm}^{-1}}

\def\Eh{E_\mathrm{h}}
\def\tr{^\mathrm{T}}

\def\np{n} %spin-1/2 particle number
\def\nb{N_\text{b}} %no. of spatial basis functions
\def\ee{\text{e}}
\def\pp{\text{p}}

\def\br{\bos{r}}
\def\bs{\bos{s}}

\def\epsi{\varepsilon}
\def\enonrel{\epsi}%{\varepsilon_0}%{\varepsilon^{(0)}}

\def\som{Supplementary Material}

\def\Sgp{^1\Sigma_\text{g}^+}
\def\Sup{^1\Sigma_\text{u}^+}
\def\Piu{^1\Pi_\text{u}}

\def\el{\text{el}}
\def\bethelog{\beta_\el}

\def\tr{^\mathrm{T}}

\definecolor{ao}{rgb}{0.0, 0.5, 0.0}

\newcolumntype{d}[1]{D{.}{.}{#1}}

\usepackage[unicode]{hyperref}
\hypersetup{
   unicode=true,          % non-Latin characters in Acrobat??s bookmarks
   plainpages=false,
   colorlinks=true,       % false: boxed links; true: colored links
   linkcolor=blue,          % color of internal links
   citecolor=blue,        % color of links to bibliography
}

\usepackage{url}

\begin{document}

\title{%
Pre-Born--Oppenheimer energies, leading-order relativistic and QED corrections 
for electronically excited states of molecular hydrogen\footnote{We dedicate this paper to Wim Ubachs on the occasion of his sixty-fifth birthday.}
}

\author{Eszter Saly} 
\author{D\'avid Ferenc} 
\author{Edit M\'atyus} 
\email{edit.matyus@ttk.elte.hu}
\affiliation{ELTE, Eötvös Loránd University, Institute of Chemistry, 
Pázmány Péter sétány 1/A, Budapest, H-1117, Hungary}

\date{\today}

\begin{abstract}
\noindent %
For rovibronic states corresponding to the $B$ and $B'\ \Sup$ electronic states of the hydrogen molecule, the pre-Born--Oppenheimer (four-particle) non-relativistic energy is converged to a 1--3 parts-per-billion relative precision. The four-particle non-relativistic energy is appended with leading-order relativistic, leading- and estimated higher-order quantum-electrodynamics corrections. The resulting term values referenced to the rovibronic ground state are obtained in an excellent agreement with the experimental results. 
Further results are reported and discussed for other rovibronic states assignable to the $C\ ^1\Pi_\text{u}$ and the $EF,GK,$ and $HH\ \Sgp$ electronic states. 
\end{abstract}

\maketitle

\section{Introduction \label{sec:intro}}
\noindent %
This work was largely motivated by the high-resolution spectroscopy experiments carried out by Wim Ubachs and his colleagues for the electronic-vibrational-rotational transitions of molecular hydrogen starting from the 1990s \cite{ReHoUb97,ReHoUbWo99,LaHoUbWo01,ReBuHoIvPeUb06,UbMe09,BaSaVeUb10,SpJuUbMe11,DiSaNiJuRoUb12,AlDrSaUbEi18,H2diss18,HoBeSaEiUbJuMe19}.

\begin{figure}
  \centering
  \includegraphics[scale=1.15]{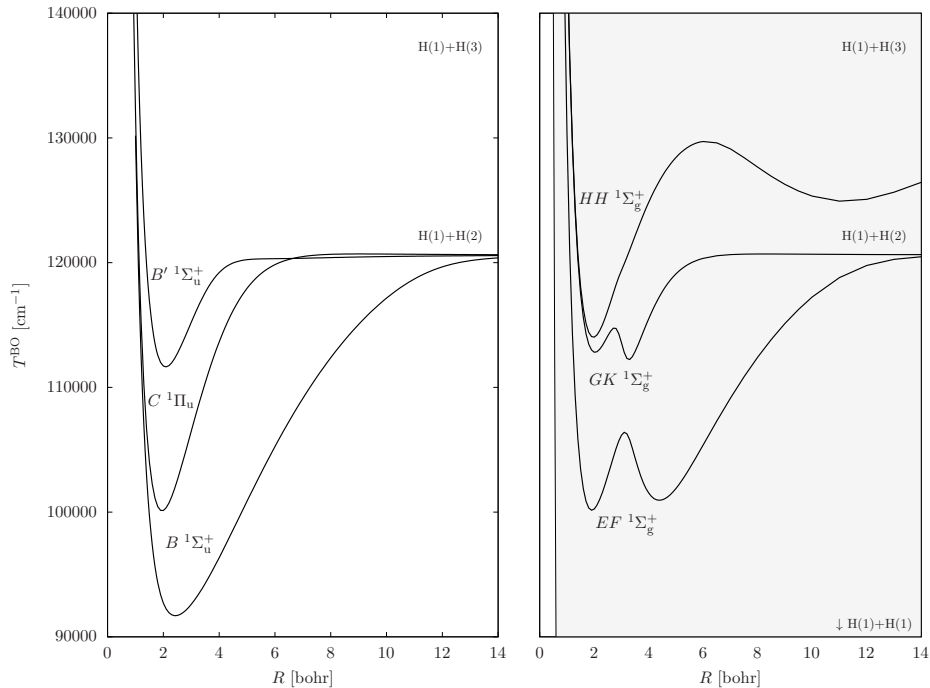}
  \caption{Rovibronic states computed in this work can be assigned to the electronic states $B,B'\ \Sup$, $C\ \Piu$, and $EF,GK,HH\ \Sgp$, for which the potential energy curves are visualized in this figure \cite{StWo02,WoSt03,SiZsPa21}. The shaded background of the gerade manifold indicates that the corresponding four-particle states are (predissociative) resonances embedded in the H(1)+H(1) continuum. The $T^{\text{BO}}$ energy is defined with respect to the minimum of the $X\ \Sgp$ electronic ground state.
  }
  \label{fig:PECs}
\end{figure}

In this work, we report computational results for the 
rovibronic states corresponding to the singlet ungerade manifold
with $N=0,1,\ldots,5$ total angular momentum quantum numbers, $p=(-1)^N$ natural parity, $v=0$ and 1 vibrational and  $B$ electronic state labels (Figure~\ref{fig:PECs}). 
Furthermore, computational results for the vibrational `ground state' of the $B'$ electronic state is also reported, which was obtained as an excited state within the 
$(N,p,S_\ee,S_\pp)=(0,+1,0,1)$
(total angular momentum, parity, total electronic and protonic spin quantum numbers) `symmetry' block \cite{Ma13}, in which the $v=0$, $B$ state is the absolute ground state.
The computed four-particle non-relativistic energies are appended with perturbative relativistic and quantum electrodynamics (QED) corrections computed in this work and using the Bethe logarithm values taken from Ref.~\cite{FeMa22bethe}.

The present work is an extension and continuation of four-particle computations carried out for the $v=0,N=0,1,2,\ldots,5,$ $EF\ \Sgp$ inner-well rotational states~\cite{FeMa19EF}, obtained in an excellent agreement with experiment \cite{DiSaNiJuRoUb12,BaSaVeUb10}.
Precise computation of electronically excited rovibronic states of the singlet gerade manifold is challenging, due to coupling to the lower-energy $\text{H}(1)+\text{H}(1)$ dissociation channel. The lowest-energy $EF$ inner-well states could be computed almost as bound states due to their very long predissociative lifetime \cite{QuDrWo90}. 
Initial results and further thoughts for other electronically excited singlet gerade states are reported at the end of this work, in relation with $EF$, $GK$, and (inner-well) $HH$ states.
Interestingly, the outer-well of $HH\ \Sgp$ is relatively weakly coupled to other nearby electronic states as it was pointed out by Wolniewicz, Ubachs, and co-workers in 1997 \cite{ReHoUb97,ReHoUbWo99}. 
In 2019, we used single-state non-adiabatic perturbation theory to improve \cite{FeMa19HH} upon the adiabatic result \cite{ReHoUb97,ReHoUbWo99}. Further improvement would be possible by multi-state non-adiabatic perturbation theory \cite{MaTe19,MaFe22nad} or perhaps by a highly efficient four-particle predissociative computation (\emph{e.g.,} by further development of Refs.~\cite{Ma13,FeMa19EF}).

\section{Theoretical framework}
\subsection{Variational solution of the four-body Schrödinger equation}
We consider H$_2=\lbrace \text{e}^-,\text{e}^-,\text{p}^+,\text{p}^+\rbrace$ as a four-particle system, and solve the translationally invariant Schrödinger equation,
\begin{align}
  H\Psi_k = E_k \Psi_k
  \label{eq:preBO}
\end{align}
with $\np+1=4$
\begin{align}
  H 
  = 
  -\frac{1}{2}\sum_{i=1}^{\np+1} \frac{1}{2m_i} \Delta_{\bos{r}_i}
  + \sum_{i=1}^{\np+1} \sum_{j>i}^{\np+1} \frac{q_i q_j}{|\bos{r}_i-\bos{r}_j|}
  +\frac{1}{2m_{1..\np+1}}\Delta_{\bos{R}_\text{CM}} \; ,
\end{align}
where Hartree atomic units are used, $m_i$ and $q_i$ are the masses and electric charges of the particles, and $\bos{R}_\text{CM}=\frac{1}{m_{1..\np+1}}\sum_{i=1}^{\np+1} m_i\bos{r}_i$ with $m_{1..\np+1}=\sum_{i=1}^{\np+1} m_i$. 

The energy (and wave function) for a selected ($k$th) eigenstate can be converged to high precision  using a non-linear variational procedure \cite{SuVaBook98,MaRe12,Ma19review}.
Each eigenstate can be characterized by the total spatial angular momentum quantum number and corresponding projection, $N,M_N$, the parity, $p$,
and the spin quantum numbers of the electrons, $S_\ee,M_{S_\ee}$, and of the protons, $S_\pp,M_{S_\pp}$. 
Hence, the wave function is approximated by a linear combination of anti-symmetrized ($\mathcal{A}$) products of $\phi_i^{[\lambda]}$ spatial and $\chi^{[\varsigma]}$ spin functions, 
\begin{align}
  \Psi^{[\lambda,\varsigma]}
  =
  \sum_{i=1}^{\nb}  
    c_{i}
    \mathcal{A} \lbrace %
      \phi_{i}^{[\lambda]}\chi^{[\varsigma]}
    \rbrace \; ,
\end{align}
where $\lambda=(N,M_N,p)$ and $\varsigma=(S_\ee,M_{S_\ee},S_\pp,M_{S_\pp})$
collect the spatial and spin quantum numbers. 
The $c_i$ linear combination coefficients are computed by solving the linear eigenvalue problem over the selected basis set. 
The $\chi^{[\varsigma]}$ spin function is the product of the singlet electron and singlet or triplet proton spin functions depending on the spatial symmetry of the computed state \cite{MaRe12,Ma13} (Table~\ref{tab:nonrel}).
For spatial functions, we use explicitly correlated Gaussian functions with the global vector representation (ECG-GVR) \cite{SuUsVa98,MaRe12} (for the precise and fully detailed definition, please consult Ref.~\citenum{MaRe12}),
\begin{align}
  \phi^{[\lambda]}(\bos{r};\bos{\alpha},\bos{u},K)
  =
  |\bos{v}|^{2K+N} Y_{NM_N}(\hat{v})
  \exp\left[%
    -\frac{1}{2}
    \sum_{i=1}^{\np+1}\sum_{j>i}^{\np+1}
      \alpha_{ij}(\bos{r}_i-\bos{r}_j)^2
  \right] \; ,
  \label{eq:ecggvr}
\end{align}
where the so-called global vector is
\begin{align}
  \bos{v} = \sum_{i=1}^{\np+1} u_i \bos{r}_i \; ,
\end{align}
and the spatial direction of $\bos{v}$, $\hat{v}=\bos{v}/|\bos{v}|$ determines the polar angles of the $Y_{NM_N}(\hat{{v}})$ spherical harmonics function. 
The non-linear parameters $\alpha_{ij}$, $u_i$, and the integer value $K=0,1,\ldots$
are optimized and fine-tuned to minimize the ($k$th) non-relativistic energy. 
For fine-tuning the real parameters ($\alpha_{ij}$ and $u_i$), we used 
the Powell method \cite{Po04}.

Analytic matrix elements for ECG-GVR functions and the non-relativistic operators have been implemented in a Fortran computer program in Ref.~\cite{MaRe12}, and this implementation has been used in practice (in double precision arithmetic in Fortran) up to ca. $K=20$ and up to $N=5-6$ \cite{MaRe12,Ma13,Ma19review,FeMa19EF}.
For the computations and the non-linear parameter optimization, the translationally invariant kinetic energy operator is expressed in terms of atoms-in-molecules internal coordinates \cite{MaRe12}.

At this point, we mention two important alternative ECG basis sets relevant for diatomic computations. 
First, ECG functions with $r_1^{2K}$-type prefactors have been efficiently used to compute diatomic molecules with $N=0$ \cite{KiAd99,KiAd00,BuAd03} and $N=1$ \cite{ShKiAd13,ShKiAd13b,KiShAd15}. For $N=0$, this basis set can also be considered as a special case of ECG-GVR with restriction of the global vector to contributions from the two protons only. Currently, our ECG-GVR implementation includes contributions from all particles, which is necessary to have a fundamentally correct total angular momentum representation beyond $N=0$. %
Second, complex ECG (CECG) functions have been proposed \cite{BuAd06,BuAd08} to efficiently describe the nodal structure of nuclear localization and excitation. In spite of the appealing simplicity of the CECG integrals, tight optimization of the non-relativistic energy with respect to the CECG (non-linear) parameters is an intricate task, due to numerical instabilities, yet CECG implementations exist for $N=0$ (and 1) \cite{BuAd06,BuAd08,Va19,MuMaRe19,BuAd20}.

\subsection{Leading-order relativistic and QED corrections\label{sec:relradcorr}}
In the non-relativistic quantum electrodynamics approach \cite{BeSabook75,Pa05FW}, the energy is written as the sum of the non-relativistic energy and correction terms for higher orders of the $\alpha$ fine-structure constant (in this work, the $\alpha$ powers are counted according to the use of Hartree atomic units),
\begin{align}
  \epsi^{(0..5)}
  =
  \enonrel 
  + 
  \alpha^2 \epsi_{\mathrm{rel}}^{(2)}
  +
  \alpha^3 \epsi_{\mathrm{qed}}^{(3)}
  +
  \alpha^4 \epsi_{\mathrm{hqed}}^{(4)}
  +
  \alpha^5 \epsi_{\mathrm{hqed}}^{(5)}    
  (+
  \ldots) \; .
\end{align}
For computing relativistic and QED corrections to  the pre-Born--Oppenheimer (preBO) energy, we consider electronic quantities (along a series of fixed nuclear configurations). The reference state is obtained by solving the electronic Schrödinger equation (for fixed nuclear configurations, $\bos{R}_1$, $\bos{R}_2$)
\begin{gather}
  H_\text{el}\phi = \enonrel \phi 
  \label{eq:Scrödinger} \\
  H_\text{el}
  =
  -\frac{1}{2}\sum_{i=1}^2 \Delta_{\bos{r}_i} 
  +\frac{1}{r_{12}}     
  -\sum_{i=1}^2 \sum_{a=1}^{2} \frac{1}{r_{ia}} 
  +\frac{1}{R} \, ,
  \label{eq:H0}
\end{gather}
with $r_{12}=|\bos{r}_{12}|=|\bos{r}_1-\bos{r}_2|$, $r_{ia}=|\bos{r}_{ia}|=|\bos{r}_i-\bos{R}_a|$, and 
$R=|\bos{R}_1-\bos{R}_2|$.
The $\phi$ electronic wave function is approximated as a linear combination of anti-symmetrized products of two-electron singlet spin and spatial functions. For spatial functions, we use floating explicitly correlated Gaussian functions (fECGs),
\begin{align}
  \varphi(\br;\bs_i,\bos{A}_i)=\exp\left[%
    -(\bos{r}-\bos{s}_i)\tr (\bos{A}_i\otimes \bos{I}_3)(\bos{r}-\bos{s}_i)
  \right]\; ,
\end{align}
which we adapt to the $\Sup$ or $\Sgp$ symmetry \cite{FeMa22bethe} of the relevant electronic state considered in this work, and optimize the $\bos{s}_i\in\mathbb{R}^{6}$ and $\bos{A}_i\in\mathbb{R}^{2\times 2}$ (symmetric, positive definite matrix) parameters based on the variational principle.

The dominant correction due to special relativity is obtained at $\alpha^2$-order as expectation value of the Breit--Pauli Hamiltonian, which, for a singlet $\phi$ electronic state, reads as
\begin{align}
\epsi_{\mathrm{rel}}^{(2)}
=
\langle %
  H_\mathrm{rel}^{(2)} 
\rangle 
=
\langle %
  \phi | H_\mathrm{rel}^{(2)} | \phi
\rangle  
\label{eq:epsitwo}
\end{align}
with
\begin{align}
H_{\mathrm{rel}}^{(2)}
=
-&\frac{1}{8} \left[ (\bos{p}_1^2)^2 + (\bos{p}_2^2)^2 \right]
+ 
\frac{\pi}{2} 
\sum_{i=1}^2\sum_{a=1}^2
  \delta(\bos{r}_{ia})   
+ \pi \delta(\bos{r}_{12}) \nonumber \\ 
&\quad\quad\quad\quad\quad-
\frac{1}{2} 
\left[%
   \frac{1}{r_{12}} \bos{p}_1\cdot\bos{p}_2
 + \frac{1}{r_{12}^3} \bos{r}_{12}(\bos{r}_{12}\cdot\bos{p}_1)\cdot\bos{p}_2
\right] \ , 
\end{align}
including the so-called mass-velocity, one-electron Darwin, two-electron Darwin plus spin-spin Fermi contact interaction, and orbit-orbit correction terms, respectively.

The spin-independent $\alpha^3$-order QED correction \cite{Ar57,sucherPhD1958} is commonly written in a compact form as \cite{BeSabook75} 
\begin{align}
  \begin{array}{@{}l}\displaystyle
  \epsi_{\rm qed}^{(3)} 
  =
  \frac{4}{3}
    \left(%
      \frac{19}{30}
      -2\ln\alpha 
      -\beta_{\rm el}
    \right)
    \sum_{i=1}^2\sum_{a=1}^2
      \langle \delta(\bos{r}_{ia}) \rangle
    +
       \left(%
         \frac{164}{15}
         +
         \frac{14}{3}\ln\alpha
       \right)
       \langle \delta(\bos{r}_{12}) \rangle 
       -\frac{7}{6\pi} P(1/r_{12}^3)
    \; ,
  \end{array}
  \label{eq:qedcorr}
\end{align}
where 
\begin{align}
  \bethelog 
  =
  -
  \frac{1}{2\pi\langle \delta\rangle}
  \langle %
    \phi | \bos{\nabla} (H_\el-\enonrel) \text{ln}[2(H_\el-\enonrel)/\Eh] \bos{\nabla} \phi
  \rangle
  \label{eq:bethelog}   
\end{align}
is the (non-relativistic) Bethe logarithm,
where $\phi$ and $\enonrel$ is the non-relativistic wave function and energy corresponding to the $H_\el$ electronic Hamiltonian, Eq.~(\ref{eq:H0}),
$\delta=\sum_{i=1}^2\sum_{a=1}^2\delta(\bos{r}_{ia})$ and 
$\bos{\nabla}=\nabla_{\bos{r}_1}+\nabla_{\bos{r}_2}$.
The so-called Araki--Sucher term \cite{Ar57,sucherPhD1958} 
(related to the retardation correction in the intermediate photon momentum range \cite{sucherPhD1958}) can be written in the usual form as
\begin{align}
  P(1/r_{12}^3)
  = 
  \lim_{\epsilon \to 0} 
  \left\langle
    \phi
    \Big| 
      \left[%
      \frac{\Theta(r_{12} - \epsilon)}{ 4\pi r_{12}^3 }
      + (\ln \epsilon + \gamma_\text{E})\delta(\bos{r}_{12}) 
      \right]
    \phi
  \right\rangle \; 
  \label{eq:Qterm}
\end{align}
with the $\Theta$ step function and the $\gamma_\text{E}$ Euler constant.

Furthermore, higher-order quantum electrodynamics corrections in this non-relativistic perturbative framework can be estimated as \cite{KoPiLaPrJePa11,PuKoCzPa19}
\begin{align}
  \epsi^{(4)}_\text{hqed,est}
  &=
  \pi\left(%
    \frac{427}{96} - 2 \ln 2
  \right)
  \sum_{i=1}^2 \sum_{a=1}^2 \langle \delta(\bos{r}_{ia}) \rangle
  \label{eq:epsifour}
  \\
  \epsi^{(5)}_\text{hqed,est}  
  &=
  -4\ln^2 \alpha 
  \sum_{i=1}^2 \sum_{a=1}^2 
    \langle \delta(\bos{r}_{ia}) \rangle \; .
  \label{eq:epsifive}    
\end{align}

Then, we compute our current best energy value for the $k$th rovibronic state as the sum of the preBO energy, $E_k$ (which we henceforth label as $E_k^{(0)}$ to emphasize that it is a non-relativistic quantity) from Eq.~(\ref{eq:preBO}), and the expectation value of the $\epsi_\kappa^{(2..5)}$ relativistic and QED corrections (for the corresponding, dominant $\kappa$th electronic state) with the $\Psi_k$ preBO wave function,
\begin{align}
  E^{(0..5)}_k
  &=
  E_k^{(0)}
  +
  \sum_{n=2}^5
    \alpha^n
    \langle %
      \Psi_k | \epsi^{(n)}_\kappa(r_{34}) | \Psi_k
    \rangle
  =
  E_k^{(0)}
  +
  \sum_{n=2}^5
    \alpha^n
    \langle %
      \epsi^{(n)}_\kappa 
    \rangle_k \; ,
    \label{eq:bestener}
\end{align}
where $r_{34}$ highlights the proton-proton distance dependence of the correction, which is integrated with the four-particle wave function to obtain the correction.

At this point, it is necessary to mention that the four-particle Breit--Pauli Hamiltonian expectation value \cite{WaYa18} and even a four-particle Bethe logarithm value \cite{PuKoCzPa19} has been reported for the rovibronic ground state ($X\ \Sgp$) of H$_2$.
The present approach is much simpler and it can be expected to give useful results, if the electronic state provides a meaningful zeroth-order approximation or in other words, if the non-adiabatic-relativistic (and QED) coupling is small. %
The overall accuracy of the present results is thus limited by the magnitude of this coupling.

\section{Computational details}
We performed pre-Born--Oppenheimer (preBO) computations for selected rovibronic states, as well as Born--Oppenheimer (BO) computations for the corresponding, dominant electronic states (along a series of nuclear configurations) using an in-house developed computer program, called QUANTEN (QUANTum mechanical computations for Electrons and atomic Nuclei) \cite{Ma19review,FeMa19HH,FeMa19EF,FeKoMa20,IrJeMaMaRoPo21,JeIrFeMa22,JeFeMa21,JeFeMa22,FeJeMa22,FeJeMa22b,MaFe22nad,FeMa22bethe}.

The preBO energies were obtained and converged within repeated variational non-linear-optimization cycles of the ECG-GVR basis set, separately carried out for every rovibronic state. Since the low-energy (ro)vibronic states corresponding to $B$ and $B'\ \Sup$ are bound states, we could use the standard variational optimization procedure applicable for a selected ($k$th, discrete) state~\cite{SuVaBook98} (Sec.~\ref{sec:ungerade}). 
Tight variational convergence of electronically excited rovibronic states from the singlet gerade manifold is more challenging due to predissociation, initial results and thoughts are collected in Sec.~\ref{sec:gerade}.

After the variational non-linear optimization procedure of the preBO energy had been sufficiently converged (including a total of 9000 basis functions optimized separately for every state), this energy was appended with relativistic and QED corrections computed as expectation value of the $\epsi^{(n)}(R)\ (n=2,3,4,5)$ correction curves with the four-particle wave function, Eq.~(\ref{eq:bestener}).

As to the leading-order relativistic electronic correction,
Wolniewicz reported values for the 
$B$ and $B''B$~$\Sup$, as well as for the $EF,\ GK,\ HH\ \Sgp$ electronic states
in Refs.~\cite{Wo95Brel,LaHoUbWo01} and \cite{Wo98EFrel}, respectively, but we could not find relativistic correction values for the $B'\ \Sup$ state. 
Regarding the leading-order quantum electrodynamics correction, the $\sum_{i,a}\langle \delta(\bos{r}_{ia})\rangle$ and $\langle \delta(\bos{r}_{12})\rangle$ values can be calculated from the one- and two-electron Darwin terms (of the relativistic correction), but we could not find any literature data for the small, though non-negligible Araki--Sucher term for these states.

To have the corrections up to $\alpha^3$-order, we computed 
the Born--Oppenheimer electronic (energies and) wave functions, and all relativistic and QED terms, Eqs.~\eqref{eq:epsitwo}, \eqref{eq:qedcorr}, \eqref{eq:Qterm}--\eqref{eq:epsifive} for the $B$ and $B'$~$\Sup$, as well as (with possible future relevance), for the
$EF,\ GK,$ and $HH\ \Sgp$ electronically excited states. 
To pinpoint the expectation value of the singular operators (mass-velocity, Dirac delta functionals, and Araki--Sucher term) with the electronic wave function described by the current fECG basis set (including 1200 functions), we used the integral transform (IT, regularization) technique originally proposed by Pachucki, Cencek, and Komasa \cite{PaCeKo05,FeKoMa20,JeIrFeMa22}. 
The relativistic and QED correction values (obtained with the IT technique) are sufficiently accurate to obtain the term correction to $1-5\cdot 10^{-9}\ \Eh$ precision \cite{FeKoMa20,JeIrFeMa22}. 
All computed data is deposited in the \som.

The computationally demanding Bethe logarithm has been recently reported by two of us for the relevant states near their equilibrium structure(s) \cite{FeMa22bethe}, and we used the value for the dominant electronic state at its (relevant) equilibrium structure (dependence on the proton-proton distance and non-adiabatic coupling was neglected).

During this work, we used the CODATA18 recommended values for constants and conversion factors \cite{codata18}.

\section{Numerical results}
\subsection{Selected rovibronic states corresponding to the singlet ungerade electronic manifold \label{sec:ungerade}}
\paragraph{$B$ and $B'$ states}
The non-relativistic preBO energies and term values computed for rovibronic states assigned (based on their energy) to the $B$ and $B'\ \Sup$ electronic states are collected in Table~\ref{tab:nonrel} and are estimated to be converged within $5\cdot 10^{-9}\ \Eh$ (0.001\ 1\ $\cm$) for the $v=0$ and within $1\cdot 10^{-8}\ \Eh$ (0.002\ 2\ $\cm$) for the $v=1$ vibrational levels.
The reported experimental uncertainty for the observed term values \cite{BaSaVeUb10} is 0.000 1--0.000 5~$\cm$. 
Table~\ref{tab:relqedcorr} collects the relativistic and quantum electrodynamics corrections to the computed term values, up to $\alpha^3$ order and estimates for the $\alpha^4$- and $\alpha^5$-order contributions. 
The final computed energy has an estimated numerical uncertainty determined by the convergence of the preBO energy, furthermore, the overall accuracy is limited by the magnitude of the non-adiabatic-relativistic coupling (which is difficult to estimate without computation).

The overall agreement of the computed results with the experimental values, within the uncertainty of both, is excellent
and represents significant improvement with respect to earlier work of Wolniewicz, Orlikowski, and Staszewska 
including non-adiabatic coupling as well as relativistic and estimated QED corrections \cite{WoOrSt06}, and the earlier, purely non-adiabatic work of Wolniewicz and Dressler \cite{WoDr92}.
We also note that Dressler and Wolniewicz \cite{DrWo95} reported a very good estimated theoretical term energy (not shown in the table) for the $B$ rovibrational ground state, obtained from a clever combination of various relevant corrections available at that time.
It is interesting to note in Table~\ref{tab:relqedcorr} that the $\alpha^3$-order corrections are (in an absolute value) larger than the $\alpha^2$-order contributions to the term energy for every reported $B$ state. This relation is reversed for the $B'$ state, though the $\alpha^3$-order correction remains significant. \\

\paragraph{$C$ state}
Finally, between the $B$ and $B'\ \Sup$ potential energy curves, we can observe the potential energy curve of the $C\ \Piu$ electronic state in Figure~\ref{fig:PECs}. We report for the rovibronic `ground-state' corresponding to $C\ \Piu$, the non-relativistic preBO energy (a variational upper bound to the exact energy) obtained and optimized as an excited state within the $(N,p,S_\ee,S_\pp)=(1,-1,0,0)$ `symmetry block' to be $E^{(0)}_{C00}=-0.712\ 254\ 357\ \Eh$. The corresponding non-relativistic term value is $T^{(0)}_{C00}=99\ 152.202\ \cm$, which is by $0.318\ \cm$ smaller, than the term energy reported by Wolniewicz, Orlikowski, and Staszewska (in the Supplementary Material) \cite{WoOrSt06}.
Wolniewicz, Orlikowski, and Staszewska write that they estimated the relativistic and radiative corrections for this state to be $\epsi^{(2..3)}_{C00}=-1.67\ \cm+0.308\ \cm=-1.362\ \cm$ \cite{WoOrSt06}. If we subtract from their term value the $\epsi^{(2..3)}_{C00}-\epsi_{X00}^{(2..3)}=-1.362\ \cm + (-2.388\ 3\ \cm + 0.736\ 1\ \cm) = 0.290\ \cm$ correction (\emph{cf.} the footnote to Table~\ref{tab:relqedcorr} for $\alpha^n\langle\epsi_X^{(n)}\rangle_{X00}\ (n=2,3)$), we find that their non-relativistic energy value (not reported separately in Ref.~\cite{WoOrSt06}) is by $0.028\ \cm$ larger, than our variational result.
Computation of the relativistic and radiative corrections to our variational four-particle energy is left for future work, and it will allow direct comparison with the experimental result \cite{BaSaVeUb10}, which is currently by $\Delta T^{(0)}_\text{C00}=-0.143\ \cm$ lower, than our non-relativistic term energy.

\begin{table}
  \caption{%
    Non-relativistic preBO energy, $E^{(0)}$ in $\Eh$, and term value, $T^{(0)}$ in $\cm$,
    for $B$ and $B'$ rovibronic states with $N=0,1,\ldots,5$ total angular momentum quantum number, natural parity, $p=(-1)^N$, and $v=0$ and 1 vibrational state labels.
    The electron spin is $S_\ee=0$, the proton spin is $S_\pp=(1+p)/2$,  \emph{i.e.,} $S_\pp=1$ proton triplet for even $N$ and $S_\pp=0$ proton singlet for odd $N$ values.
    \label{tab:nonrel}
  }%
  \begin{tabular}{@{}c@{\ \ }r@{\ \ }r@{\ \ }r@{}}
    \hline\hline\\[-0.35cm]  
    $N$ & 
    $E^{(0)}$ & 
    $T^{(0)}\ ^\text{a}$ & 
    $\Delta T^{(0)}_{\text{o-c}}\ ^\text{b}$ 
    \\
    \hline\\[-0.35cm]      
    \multicolumn{4}{l}{$B,v=0$:} \\
    0	&	$-$0.753 027 186	&	90 203.600 5	&	$-$0.100 7 \\ 
    1	&	$-$0.752 850 233	&	90 242.437 1	&	$-$0.099 0 \\
    2	&	$-$0.752 498 025	&	90 319.737 9	&	$-$0.095 2 \\
    3	&	$-$0.751 973 903	&	90 434.769 4	&	$-$0.089 8 \\
    4	&	$-$0.751 282 748	&	90 586.460 3	&	$-$0.082 8 \\
    5	&	$-$0.750 430 841	&	90 773.432 4	&	$-$0.073 1 \\
    \hline\\[-0.35cm]
    \multicolumn{4}{l}{$B,v=1$:} \\
    0	&	$-$0.747 020 699	&	91 521.871 9	&	$-$0.047 2 \\
    1	&	$-$0.746 852 576	&	91 558.770 8	&	$-$0.045 9 \\
    2	&	$-$0.746 517 802	&	91 632.245 2	&	$-$0.043 2 \\
    3	&	$-$0.746 019 274	&	91 741.659 3	&	$-$0.036 8 \\
    4	&	$-$0.745 361 230	&	91 886.083 4	&	$-$0.030 4 \\
    5	&	$-$0.744 549 144	&	92 064.315 7	&	$-$0.021 9  \\
    \hline\\[-0.35cm]
    \multicolumn{4}{l}{$B',v=0$:} \\
    0	&	$-$0.660 648 744	&	110 478.325 0	&	0.214 9 \\
    \hline\hline\\[-0.35cm]  
  \end{tabular}
  \begin{flushleft}
  $^\text{a}$~%
  The non-relativistic term value was calculated as $T^{(0)}=E^{(0)}-E^{(0)}_{X00}$ using the $E^{(0)}_{X00}=-1.164\ 025\ 030\ 9\ \Eh$ non-relativistic energy for the rovibronic ground state ($X00$) \cite{PuKoCzPa19}.
  $^\text{b}$~%
    $\Delta T^{(0)}_{\text{o-c}}=T_\text{o}-T^{(0)}$
    deviation of the $T_\text{o}$ observed \cite{BaSaVeUb10} and $T^{(0)}$ computed (this work) term values.
  \end{flushleft}
\end{table}

\begin{table}
  \caption{%
    Perturbative relativistic and QED corrections computed to the non-relativistic preBO term energies (Table~\ref{tab:nonrel}), in $\cm$. The estimated numerical uncertainty of the $T^{(0..5)}$ total term value is $0.002\ \cm$ and $0.005\ \cm$ 
    for the $v=0$ and $v=1$ vibrational levels, respectively. 
    \label{tab:relqedcorr}
  }%
  \begin{tabular}{@{}c@{\ \ }c@{\ \ }r@{\ \ }r@{\ \ }r@{\ \ }r@{\ \ }r@{\ \ }c@{\ \ }c @{}}
    \hline\hline\\[-0.35cm]  
    $N$ & 
    $\delta T^{(2)}_\text{rel}\ ^\text{a}$ &
    $\delta T^{(3)}_\text{qed}\ ^\text{a}$ & 
    $\delta T^{(4..5)}_\text{hqed}\ ^\text{a}$ & 
    $\delta T^{(0..5)}\ ^\text{a}$ &
    $T^{(0..5)}\ ^\text{a}$ &
    $\Delta T^{(0..5)}_\text{o-c}\ ^\text{b}$ &
    $\Delta T^{(0..3)}_\text{o-c}\ ^\text{c}$ \cite{WoOrSt06} &
    $\Delta T^{(0)}_\text{o-c}$ \cite{WoDr92} 
    \\
    \hline\\[-0.35cm]      
    \multicolumn{4}{l}{$B\ \Sup,v=0$:} \\
    0	&	0.299 6	&	$-$0.398 4	&	$-$0.003 0	&	$-$0.101 8	& 90 203.498 7 &	0.001 1	& $-$0.27 & $-$1.18 \\ %
    1	&	0.301 7	&	$-$0.398 6	&	$-$0.003 0	&	$-$0.099 9	& 90 242.337 2 &	0.000 9	& $-$0.28 & \\
    2	&	0.306 0	&	$-$0.399 1	&	$-$0.003 0	&	$-$0.096 1	& 90 319.641 7 &	0.000 9	& $-$0.30 & \\
    3	&	0.312 3	&	$-$0.399 8	&	$-$0.003 0	&	$-$0.090 5	& 90 434.678 9 &	0.000 7	& $-$0.32 & \\
    4	&	0.320 7	&	$-$0.400 7	&	$-$0.003 0	&	$-$0.083 0	& 90 586.377 2 &	0.000 3	& $-$0.36 & \\
    5	&	0.330 9	&	$-$0.401 8	&	$-$0.003 0	&	$-$0.073 9	& 90 773.358 5 &	0.000 8	& $-$0.40 & \\
    \hline\\[-0.35cm]
    \multicolumn{4}{l}{$B\ \Sup,v=1$:} \\
    0	&	0.358 9	&	$-$0.400 2	&	$-$0.003 0	&	$-$0.044 3	& 91 521.827 6 &	$-$0.002 8	& $-$0.53 & $-$1.16 \\
    1	&	0.360 8	&	$-$0.400 4	&	$-$0.003 0	&	$-$0.042 6	& 91 558.728 2 &	$-$0.003 3	& $-$0.53 & \\
    2	&	0.364 6	&	$-$0.400 8	&	$-$0.003 0	&	$-$0.039 2	& 91 632.205 9 &	$-$0.003 9	& $-$0.54 & \\
    3	&	0.370 3	&	$-$0.401 4	&	$-$0.003 0	&	$-$0.034 2	& 91 741.625 2 &	$-$0.002 7	& $-$0.57 & \\
    4	&	0.377 8	&	$-$0.402 3	&	$-$0.003 0	&	$-$0.027 5	& 91 886.055 9 &	$-$0.002 9	& $-$0.60 & \\
    5	&	0.386 9	&	$-$0.403 2	&	$-$0.003 0	&	$-$0.019 4	& 92 064.296 3 &	$-$0.002 5	& $-$0.64 & \\
    \hline\\[-0.35cm]
    \multicolumn{4}{l}{$B'\ \Sup,v=0$:} \\
    0	&	0.622 9	&	$-$0.403 6	&	$-$0.003 0	&	   0.216 3	& 110 478.541 3 &	$-$0.001 4	& $-$0.31 & $-$1.87 \\
    \hline\hline\\[-0.35cm]  
  \end{tabular}
  \begin{flushleft}
  $^\text{a}$~%
    Corrections and total term energy, $T=E^{(0..5)}-E^{(0..5)}_{X00}$, reported with respect to the rovibronic ground state ($X00$).  
    $E^{(0..5)}$ was computed in this work, Eq.~\eqref{eq:bestener}.
    The $E^{(0..5)}_{X00}$ value for the rovibronic ground state was obtained as the sum of the $E^{(0)}_{X00}$ non-relativistic energy (see footnote to Table~\ref{tab:nonrel}), the $\alpha^2\langle \epsi^{(2)}_X\rangle_{X00}=-2.388\ 3~\cm$ relativistic, 
    $\alpha^3 \langle\epsi^{(3)}_X\rangle_{X00}=0.736\ 1~\cm$ leading-order QED, 
    and $\alpha^4 \langle\epsi^{(4)}_X\rangle_{X00}+\alpha^5\langle\epsi^{(5)}_X\rangle_{X00}=0.005\ 3~\cm$ estimated higher-order QED corrections calculated with data
    compiled from \cite{PuKoPa17}. \\
  $^\text{b}$~%
    $\Delta T^{(0..5)}_{\text{o-c}}=T_\text{o}-T^{(0..5)}$
    deviation of the $T_\text{o}$ observed \cite{BaSaVeUb10} and $T^{(0..5)}$ computed (this work) term values. \\
  $^\text{c}$~%
    The differences are obtained from the experimental term energies of Ref.~\cite{BaSaVeUb10} and the computed term energies listed in the Supplementary Material of Ref.~\cite{WoOrSt06}.
  \end{flushleft}
\end{table}

\clearpage

\subsection{Initial results regarding rovibronic states of the singlet gerade electronic manifold \label{sec:gerade}}
We report initial results and observations regarding selected rovibronic states from the singlet gerade ($EF,GK,HH$) manifold. 
The electronic expectation values for the relativistic and QED corrections (most importantly, the Araki--Sucher term) are deposited in the \som, and the Bethe logarithm for these states near the equilibrium structures are available from Ref.~\cite{FeMa22bethe}.
For computation of the preBO energy, an initial basis parameter set was compiled from earlier bound-state computations \cite{MaRe12,Ma13,Ma19review,FeMa19EF}. Then, the basis representation of the (long-lived, but predissociative) rovibronic states was optimized in a stabilization-like procedure~\cite{FeMa19EF}. \\

\paragraph{Inner-well $HH$ state}
Regarding the inner well of the $HH\ \Sgp$ electronic state, we could converge the preBO energy for the rotational-vibrational `ground' state ($v=0$) as excited state within the $(N,p,S_\ee,S_\pp)=(0,+1,0,0)$ symmetry block,
it is $E^{(0)}_{HH00}=-0.649\ 353\ 094(5)\ \Eh$ (with an estimated $5\ \text{n}\Eh$ numerical uncertainty), and the non-relativistic term energy is $T^{(0)}_{HH,00}=112\ 957.434(1)\ \cm$ (referenced to the non-relativistic rovibronic ground-state energy, see footnote to Table~\ref{tab:nonrel}). This non-relativistic term value differs from experiment \cite{BaSaVeUb10} by $\Delta T^{(0)}_\text{o-c}=0.125\ \cm$, which is less than the $0.89\ \cm$ deviation of the non-relativistic, non-adiabatic result reported by Yu and Dressler \cite{YuDr94}.
For the relativistic and QED corrections, we could not pinpoint the numerically stable result using the procedure of Sec.~\ref{sec:relradcorr} and Eq.~\eqref{eq:bestener}, and direct computation of the four-particle relativistic corrections appears to be necessary. \\

\paragraph{$GK$ states}
Regarding the optimization of the lowest-energy vibrational states corresponding to the $GK\ \Sgp$ electronic state, we could observe the `$G0$' ($v=1$ for $GK$) and also the `$K0$' ($v=0$ for $GK$) states, but their non-relativistic energy could be obtained only with a ca. $1\ \mu\Eh$ uncertainty in a stabilization-like preBO computation \cite{FeMa19EF}. Thus, our current best estimated four-particle energies are
$E^{(0)}_{K00}=-0.655\ 406(5)\ \Eh$ and $E^{(0)}_{G00}=-0.654\ 568(5)\ \Eh$. 
The corresponding term values are 
$T^{(0)}_{K00}=111\ 629(1)\ \cm$ and $T^{(0)}_{G00}=111\ 813(1)\ \cm$ 
with $-0.2$ and $-0.3\ \cm$ deviation from experiment \cite{BaSaVeUb10,HoBeMe18}, which is smaller than the deviation of the non-adiabatic value, $3.38$ and $2.635\ \cm$, reported by Yu and Dressler \cite{YuDr94}. 
Further improvement of the optimization procedure, perhaps in combination with optimization of the complex-energy of the complex-coordinate rotated Hamiltonian \cite{MaRe12,BaZhYaSh21} could be used to pinpoint the four-particle energy (as a resonance state). \\

\paragraph{$EF$ states}
As to rovibronic states corresponding to the $EF\ \Sgp$ electronic state, rotational excitations of the vibrational ground state ($E0N$) of the `$E$' inner well have been reported to high precision including relativistic, QED, and higher-order QED corrections in an excellent agreement \cite{FeMa19EF} with the (more accurate) experimental results. 
In this work, we could also converge the preBO energy of the first vibrationally excited state assigned to the inner well ($E10$, but $v=3$ for $EF$), it is $E^{(0)}_{E10}=-0.701\ 581\ 805(5)\ \Eh$ and $T^{(0)}_{E10}=101\ 494.556(1)\ \cm$ with $\Delta T^{(0)}_{\text{o-c},E10}=0.188\ \cm$ experiment-theory deviation, which is smaller than the $0.746\ \cm$ value from the non-adiabatic computations of Yu and Dressler \cite{YuDr94}. In further work, relativistic and QED corrections to this state should be obtained in direct, four-particle computations.

Regarding the `$F$' outer-well states, our current best four-particle energy for the rovibrational ground state ($F00$, but $v=1$ for $EF$) is $E^{(0)}_{F00}=-0.711\ 287(5)\ \Eh$, $T^{(0)}_{F00}=99\ 365(1)\ \cm$, which differs from the experimental value by $-0.7\ \cm$. This deviation is larger than the $0.14\ \cm$ difference of the non-adiabatic result of Yu and Dressler \cite{YuDr94}. In our procedure, the non-linear optimization of this state is stable, but currently inefficient. To speed up the computation, it will be necessary to find a better starting basis set (the current starting parameterization was compiled from basis functions optimized for rovibronic states with ca. 2~bohr proton-proton expectation value, which is much smaller than the proton-proton distance characteristic for $F$ states).

\section{Summary and conclusion \label{sec:summary}}
Rovibronic states from the singlet ungerade ($B, B',C$) and the singlet gerade ($EF,GK,HH$) manifolds of molecular hydrogen have been computed by solving the electron-proton four-particle Schrödinger equation. 
For the lowest-energy rovibronic $B$ and $B'$ states, an excellent agreement of the rovibronic term value (referenced to the absolute ground state) is obtained with respect to high-resolution experiment after inclusion of the relativistic, leading-order and estimated higher-order quantum-electrodynamics corrections. 
For the other states, the current status and necessary further developments are discussed.

\noindent %\vspace{0.5cm}
\begin{acknowledgments}
\noindent Financial support of the European Research Council through a Starting Grant (No.~851421) is gratefully acknowledged. DF thanks a doctoral scholarship from the ÚNKP-22-4 New National Excellence Program of the Ministry for Innovation and Technology from the source of the National Research, Development and Innovation Fund (ÚNKP-22-4-I-ELTE-51). 
\end{acknowledgments}

\clearpage
%\bibliography{mybib}

\begin{thebibliography}{62}%
\makeatletter
\providecommand \@ifxundefined [1]{%
 \@ifx{#1\undefined}
}%
\providecommand \@ifnum [1]{%
 \ifnum #1\expandafter \@firstoftwo
 \else \expandafter \@secondoftwo
 \fi
}%
\providecommand \@ifx [1]{%
 \ifx #1\expandafter \@firstoftwo
 \else \expandafter \@secondoftwo
 \fi
}%
\providecommand \natexlab [1]{#1}%
\providecommand \enquote  [1]{``#1''}%
\providecommand \bibnamefont  [1]{#1}%
\providecommand \bibfnamefont [1]{#1}%
\providecommand \citenamefont [1]{#1}%
\providecommand \href@noop [0]{\@secondoftwo}%
\providecommand \href [0]{\begingroup \@sanitize@url \@href}%
\providecommand \@href[1]{\@@startlink{#1}\@@href}%
\providecommand \@@href[1]{\endgroup#1\@@endlink}%
\providecommand \@sanitize@url [0]{\catcode `\\12\catcode `\$12\catcode
  `\&12\catcode `\#12\catcode `\^12\catcode `\_12\catcode `\%12\relax}%
\providecommand \@@startlink[1]{}%
\providecommand \@@endlink[0]{}%
\providecommand \url  [0]{\begingroup\@sanitize@url \@url }%
\providecommand \@url [1]{\endgroup\@href {#1}{\urlprefix }}%
\providecommand \urlprefix  [0]{URL }%
\providecommand \Eprint [0]{\href }%
\providecommand \doibase [0]{http://dx.doi.org/}%
\providecommand \selectlanguage [0]{\@gobble}%
\providecommand \bibinfo  [0]{\@secondoftwo}%
\providecommand \bibfield  [0]{\@secondoftwo}%
\providecommand \translation [1]{[#1]}%
\providecommand \BibitemOpen [0]{}%
\providecommand \bibitemStop [0]{}%
\providecommand \bibitemNoStop [0]{.\EOS\space}%
\providecommand \EOS [0]{\spacefactor3000\relax}%
\providecommand \BibitemShut  [1]{\csname bibitem#1\endcsname}%
\let\auto@bib@innerbib\@empty
%</preamble>
\bibitem [{\citenamefont {Reinhold}\ \emph {et~al.}(1997)\citenamefont
  {Reinhold}, \citenamefont {Hogervorst},\ and\ \citenamefont
  {Ubachs}}]{ReHoUb97}%
  \BibitemOpen
  \bibfield  {author} {\bibinfo {author} {\bibfnamefont {E.}~\bibnamefont
  {Reinhold}}, \bibinfo {author} {\bibfnamefont {W.}~\bibnamefont
  {Hogervorst}}, \ and\ \bibinfo {author} {\bibfnamefont {W.}~\bibnamefont
  {Ubachs}},\ }\href {\doibase 10.1103/PhysRevLett.78.2543} {\bibfield
  {journal} {\bibinfo  {journal} {Phys. Rev. Lett.}\ }\textbf {\bibinfo
  {volume} {78}},\ \bibinfo {pages} {2543} (\bibinfo {year}
  {1997})}\BibitemShut {NoStop}%
\bibitem [{\citenamefont {Reinhold}\ \emph {et~al.}(1999)\citenamefont
  {Reinhold}, \citenamefont {Hogervorst}, \citenamefont {Ubachs},\ and\
  \citenamefont {Wolniewicz}}]{ReHoUbWo99}%
  \BibitemOpen
  \bibfield  {author} {\bibinfo {author} {\bibfnamefont {E.}~\bibnamefont
  {Reinhold}}, \bibinfo {author} {\bibfnamefont {W.}~\bibnamefont
  {Hogervorst}}, \bibinfo {author} {\bibfnamefont {W.}~\bibnamefont {Ubachs}},
  \ and\ \bibinfo {author} {\bibfnamefont {L.}~\bibnamefont {Wolniewicz}},\
  }\href {\doibase 10.1103/PhysRevA.60.1258} {\bibfield  {journal} {\bibinfo
  {journal} {Phys. Rev. A}\ }\textbf {\bibinfo {volume} {60}},\ \bibinfo
  {pages} {1258} (\bibinfo {year} {1999})}\BibitemShut {NoStop}%
\bibitem [{\citenamefont {de~Lange}\ \emph {et~al.}(2001)\citenamefont
  {de~Lange}, \citenamefont {Hogervorst}, \citenamefont {Ubachs},\ and\
  \citenamefont {Wolniewicz}}]{LaHoUbWo01}%
  \BibitemOpen
  \bibfield  {author} {\bibinfo {author} {\bibfnamefont {A.}~\bibnamefont
  {de~Lange}}, \bibinfo {author} {\bibfnamefont {W.}~\bibnamefont
  {Hogervorst}}, \bibinfo {author} {\bibfnamefont {W.}~\bibnamefont {Ubachs}},
  \ and\ \bibinfo {author} {\bibfnamefont {L.}~\bibnamefont {Wolniewicz}},\
  }\href {\doibase 10.1103/PhysRevLett.86.2988} {\bibfield  {journal} {\bibinfo
   {journal} {Phys. Rev. Lett.}\ }\textbf {\bibinfo {volume} {86}},\ \bibinfo
  {pages} {2988} (\bibinfo {year} {2001})}\BibitemShut {NoStop}%
\bibitem [{\citenamefont {Reinhold}\ \emph {et~al.}(2006)\citenamefont
  {Reinhold}, \citenamefont {Buning}, \citenamefont {Hollenstein},
  \citenamefont {Ivanchik}, \citenamefont {Petitjean},\ and\ \citenamefont
  {Ubachs}}]{ReBuHoIvPeUb06}%
  \BibitemOpen
  \bibfield  {author} {\bibinfo {author} {\bibfnamefont {E.}~\bibnamefont
  {Reinhold}}, \bibinfo {author} {\bibfnamefont {R.}~\bibnamefont {Buning}},
  \bibinfo {author} {\bibfnamefont {U.}~\bibnamefont {Hollenstein}}, \bibinfo
  {author} {\bibfnamefont {A.}~\bibnamefont {Ivanchik}}, \bibinfo {author}
  {\bibfnamefont {P.}~\bibnamefont {Petitjean}}, \ and\ \bibinfo {author}
  {\bibfnamefont {W.}~\bibnamefont {Ubachs}},\ }\href {\doibase
  10.1103/PhysRevLett.96.151101} {\bibfield  {journal} {\bibinfo  {journal}
  {Phys. Rev. Lett.}\ }\textbf {\bibinfo {volume} {96}},\ \bibinfo {pages}
  {151101} (\bibinfo {year} {2006})}\BibitemShut {NoStop}%
\bibitem [{\citenamefont {Liu}\ \emph {et~al.}(2009)\citenamefont {Liu},
  \citenamefont {Salumbides}, \citenamefont {Hollenstein}, \citenamefont
  {Koelemeij}, \citenamefont {Eikema}, \citenamefont {Ubachs},\ and\
  \citenamefont {Merkt}}]{UbMe09}%
  \BibitemOpen
  \bibfield  {author} {\bibinfo {author} {\bibfnamefont {J.}~\bibnamefont
  {Liu}}, \bibinfo {author} {\bibfnamefont {E.~J.}\ \bibnamefont {Salumbides}},
  \bibinfo {author} {\bibfnamefont {U.}~\bibnamefont {Hollenstein}}, \bibinfo
  {author} {\bibfnamefont {J.~C.~J.}\ \bibnamefont {Koelemeij}}, \bibinfo
  {author} {\bibfnamefont {K.~S.~E.}\ \bibnamefont {Eikema}}, \bibinfo {author}
  {\bibfnamefont {W.}~\bibnamefont {Ubachs}}, \ and\ \bibinfo {author}
  {\bibfnamefont {F.}~\bibnamefont {Merkt}},\ }\href {\doibase
  10.1063/1.3120443} {\bibfield  {journal} {\bibinfo  {journal} {J. Chem.
  Phys.}\ }\textbf {\bibinfo {volume} {130}},\ \bibinfo {pages} {174306}
  (\bibinfo {year} {2009})}\BibitemShut {NoStop}%
\bibitem [{\citenamefont {Bailly}\ \emph {et~al.}(2010)\citenamefont {Bailly},
  \citenamefont {Salumbides}, \citenamefont {Vervloet},\ and\ \citenamefont
  {Ubachs}}]{BaSaVeUb10}%
  \BibitemOpen
  \bibfield  {author} {\bibinfo {author} {\bibfnamefont {D.}~\bibnamefont
  {Bailly}}, \bibinfo {author} {\bibfnamefont {E.~J.}\ \bibnamefont
  {Salumbides}}, \bibinfo {author} {\bibfnamefont {M.}~\bibnamefont
  {Vervloet}}, \ and\ \bibinfo {author} {\bibfnamefont {W.}~\bibnamefont
  {Ubachs}},\ }\href {\doibase 10.1080/00268970903413350} {\bibfield  {journal}
  {\bibinfo  {journal} {Mol. Phys.}\ }\textbf {\bibinfo {volume} {108}},\
  \bibinfo {pages} {827} (\bibinfo {year} {2010})}\BibitemShut {NoStop}%
\bibitem [{\citenamefont {Sprecher}\ \emph {et~al.}(2011)\citenamefont
  {Sprecher}, \citenamefont {Jungen}, \citenamefont {Ubachs},\ and\
  \citenamefont {Merkt}}]{SpJuUbMe11}%
  \BibitemOpen
  \bibfield  {author} {\bibinfo {author} {\bibfnamefont {D.}~\bibnamefont
  {Sprecher}}, \bibinfo {author} {\bibfnamefont {C.}~\bibnamefont {Jungen}},
  \bibinfo {author} {\bibfnamefont {W.}~\bibnamefont {Ubachs}}, \ and\ \bibinfo
  {author} {\bibfnamefont {F.}~\bibnamefont {Merkt}},\ }\href {\doibase
  10.1039/C0FD00035C} {\bibfield  {journal} {\bibinfo  {journal} {Farad.
  Discuss.}\ }\textbf {\bibinfo {volume} {150}},\ \bibinfo {pages} {51}
  (\bibinfo {year} {2011})}\BibitemShut {NoStop}%
\bibitem [{\citenamefont {Dickenson}\ \emph {et~al.}(2012)\citenamefont
  {Dickenson}, \citenamefont {Salumbides}, \citenamefont {Niu}, \citenamefont
  {Jungen}, \citenamefont {Ross},\ and\ \citenamefont
  {Ubachs}}]{DiSaNiJuRoUb12}%
  \BibitemOpen
  \bibfield  {author} {\bibinfo {author} {\bibfnamefont {G.~D.}\ \bibnamefont
  {Dickenson}}, \bibinfo {author} {\bibfnamefont {E.~J.}\ \bibnamefont
  {Salumbides}}, \bibinfo {author} {\bibfnamefont {M.}~\bibnamefont {Niu}},
  \bibinfo {author} {\bibfnamefont {C.}~\bibnamefont {Jungen}}, \bibinfo
  {author} {\bibfnamefont {S.~C.}\ \bibnamefont {Ross}}, \ and\ \bibinfo
  {author} {\bibfnamefont {W.}~\bibnamefont {Ubachs}},\ }\href {\doibase
  10.1103/PhysRevA.86.032502} {\bibfield  {journal} {\bibinfo  {journal} {Phys.
  Rev. A}\ }\textbf {\bibinfo {volume} {86}},\ \bibinfo {pages} {032502}
  (\bibinfo {year} {2012})}\BibitemShut {NoStop}%
\bibitem [{\citenamefont {Altmann}\ \emph {et~al.}(2018)\citenamefont
  {Altmann}, \citenamefont {Dreissen}, \citenamefont {Salumbides},
  \citenamefont {Ubachs},\ and\ \citenamefont {Eikema}}]{AlDrSaUbEi18}%
  \BibitemOpen
  \bibfield  {author} {\bibinfo {author} {\bibfnamefont {R.~K.}\ \bibnamefont
  {Altmann}}, \bibinfo {author} {\bibfnamefont {L.~S.}\ \bibnamefont
  {Dreissen}}, \bibinfo {author} {\bibfnamefont {E.~J.}\ \bibnamefont
  {Salumbides}}, \bibinfo {author} {\bibfnamefont {W.}~\bibnamefont {Ubachs}},
  \ and\ \bibinfo {author} {\bibfnamefont {K.~S.~E.}\ \bibnamefont {Eikema}},\
  }\href {\doibase 10.1103/PhysRevLett.120.043204} {\bibfield  {journal}
  {\bibinfo  {journal} {Phys. Rev. Lett.}\ }\textbf {\bibinfo {volume} {120}},\
  \bibinfo {pages} {043204} (\bibinfo {year} {2018})}\BibitemShut {NoStop}%
\bibitem [{\citenamefont {Cheng}\ \emph {et~al.}(2018)\citenamefont {Cheng},
  \citenamefont {Hussels}, \citenamefont {Niu}, \citenamefont {Bethlem},
  \citenamefont {Eikema}, \citenamefont {Salumbides}, \citenamefont {Ubachs},
  \citenamefont {Beyer}, \citenamefont {H\"olsch}, \citenamefont {Agner},
  \citenamefont {Merkt}, \citenamefont {Tao}, \citenamefont {Hu},\ and\
  \citenamefont {Jungen}}]{H2diss18}%
  \BibitemOpen
  \bibfield  {author} {\bibinfo {author} {\bibfnamefont {C.-F.}\ \bibnamefont
  {Cheng}}, \bibinfo {author} {\bibfnamefont {J.}~\bibnamefont {Hussels}},
  \bibinfo {author} {\bibfnamefont {M.}~\bibnamefont {Niu}}, \bibinfo {author}
  {\bibfnamefont {H.~L.}\ \bibnamefont {Bethlem}}, \bibinfo {author}
  {\bibfnamefont {K.~S.~E.}\ \bibnamefont {Eikema}}, \bibinfo {author}
  {\bibfnamefont {E.~J.}\ \bibnamefont {Salumbides}}, \bibinfo {author}
  {\bibfnamefont {W.}~\bibnamefont {Ubachs}}, \bibinfo {author} {\bibfnamefont
  {M.}~\bibnamefont {Beyer}}, \bibinfo {author} {\bibfnamefont
  {N.}~\bibnamefont {H\"olsch}}, \bibinfo {author} {\bibfnamefont {J.~A.}\
  \bibnamefont {Agner}}, \bibinfo {author} {\bibfnamefont {F.}~\bibnamefont
  {Merkt}}, \bibinfo {author} {\bibfnamefont {L.-G.}\ \bibnamefont {Tao}},
  \bibinfo {author} {\bibfnamefont {S.-M.}\ \bibnamefont {Hu}}, \ and\ \bibinfo
  {author} {\bibfnamefont {C.}~\bibnamefont {Jungen}},\ }\href {\doibase
  10.1103/PhysRevLett.121.013001} {\bibfield  {journal} {\bibinfo  {journal}
  {Phys. Rev. Lett.}\ }\textbf {\bibinfo {volume} {121}},\ \bibinfo {pages}
  {013001} (\bibinfo {year} {2018})}\BibitemShut {NoStop}%
\bibitem [{\citenamefont {H\"olsch}\ \emph {et~al.}(2019)\citenamefont
  {H\"olsch}, \citenamefont {Beyer}, \citenamefont {Salumbides}, \citenamefont
  {Eikema}, \citenamefont {Ubachs}, \citenamefont {Jungen},\ and\ \citenamefont
  {Merkt}}]{HoBeSaEiUbJuMe19}%
  \BibitemOpen
  \bibfield  {author} {\bibinfo {author} {\bibfnamefont {N.}~\bibnamefont
  {H\"olsch}}, \bibinfo {author} {\bibfnamefont {M.}~\bibnamefont {Beyer}},
  \bibinfo {author} {\bibfnamefont {E.~J.}\ \bibnamefont {Salumbides}},
  \bibinfo {author} {\bibfnamefont {K.~S.}\ \bibnamefont {Eikema}}, \bibinfo
  {author} {\bibfnamefont {W.}~\bibnamefont {Ubachs}}, \bibinfo {author}
  {\bibfnamefont {C.}~\bibnamefont {Jungen}}, \ and\ \bibinfo {author}
  {\bibfnamefont {F.}~\bibnamefont {Merkt}},\ }\href {\doibase
  10.1103/PhysRevLett.122.103002} {\bibfield  {journal} {\bibinfo  {journal}
  {Phys. Rev. Lett.}\ }\textbf {\bibinfo {volume} {122}},\ \bibinfo {pages}
  {103002} (\bibinfo {year} {2019})}\BibitemShut {NoStop}%
\bibitem [{\citenamefont {Mátyus}(2013)}]{Ma13}%
  \BibitemOpen
  \bibfield  {author} {\bibinfo {author} {\bibfnamefont {E.}~\bibnamefont
  {Mátyus}},\ }\href {\doibase 10.1021/jp4010696} {\bibfield  {journal}
  {\bibinfo  {journal} {J. Phys. Chem. A}\ }\textbf {\bibinfo {volume} {117}},\
  \bibinfo {pages} {7195} (\bibinfo {year} {2013})}\BibitemShut {NoStop}%
\bibitem [{\citenamefont {Ferenc}\ and\ \citenamefont
  {Mátyus}(2023)}]{FeMa22bethe}%
  \BibitemOpen
  \bibfield  {author} {\bibinfo {author} {\bibfnamefont {D.}~\bibnamefont
  {Ferenc}}\ and\ \bibinfo {author} {\bibfnamefont {E.}~\bibnamefont
  {Mátyus}},\ }\href {\doibase 10.1021/acs.jpca.2c05790} {\bibfield  {journal}
  {\bibinfo  {journal} {J. Phys. Chem. A}\ } (\bibinfo {year} {2023}),\
  10.1021/acs.jpca.2c05790}\BibitemShut {NoStop}%
\bibitem [{\citenamefont {Ferenc}\ and\ \citenamefont
  {M\'atyus}(2019)}]{FeMa19EF}%
  \BibitemOpen
  \bibfield  {author} {\bibinfo {author} {\bibfnamefont {D.}~\bibnamefont
  {Ferenc}}\ and\ \bibinfo {author} {\bibfnamefont {E.}~\bibnamefont
  {M\'atyus}},\ }\href {\doibase 10.1103/PhysRevA.100.020501} {\bibfield
  {journal} {\bibinfo  {journal} {Phys. Rev. A}\ }\textbf {\bibinfo {volume}
  {100}},\ \bibinfo {pages} {020501(R)} (\bibinfo {year} {2019})}\BibitemShut
  {NoStop}%
\bibitem [{\citenamefont {Quadrelli}\ \emph {et~al.}(1990)\citenamefont
  {Quadrelli}, \citenamefont {Dressler},\ and\ \citenamefont
  {Wolniewicz}}]{QuDrWo90}%
  \BibitemOpen
  \bibfield  {author} {\bibinfo {author} {\bibfnamefont {P.}~\bibnamefont
  {Quadrelli}}, \bibinfo {author} {\bibfnamefont {K.}~\bibnamefont {Dressler}},
  \ and\ \bibinfo {author} {\bibfnamefont {L.}~\bibnamefont {Wolniewicz}},\
  }\href {\doibase 10.1063/1.458633} {\bibfield  {journal} {\bibinfo  {journal}
  {J. Chem. Phys.}\ }\textbf {\bibinfo {volume} {93}},\ \bibinfo {pages} {4958}
  (\bibinfo {year} {1990})}\BibitemShut {NoStop}%
\bibitem [{\citenamefont {Ferenc}\ and\ \citenamefont
  {Mátyus}(2019)}]{FeMa19HH}%
  \BibitemOpen
  \bibfield  {author} {\bibinfo {author} {\bibfnamefont {D.}~\bibnamefont
  {Ferenc}}\ and\ \bibinfo {author} {\bibfnamefont {E.}~\bibnamefont
  {Mátyus}},\ }\href {\doibase 10.1063/1.5109964} {\bibfield  {journal}
  {\bibinfo  {journal} {J. Chem. Phys.}\ }\textbf {\bibinfo {volume} {151}},\
  \bibinfo {pages} {094101} (\bibinfo {year} {2019})}\BibitemShut {NoStop}%
\bibitem [{\citenamefont {M\'atyus}\ and\ \citenamefont
  {Teufel}(2019)}]{MaTe19}%
  \BibitemOpen
  \bibfield  {author} {\bibinfo {author} {\bibfnamefont {E.}~\bibnamefont
  {M\'atyus}}\ and\ \bibinfo {author} {\bibfnamefont {S.}~\bibnamefont
  {Teufel}},\ }\href {\doibase 10.1063/1.5097899} {\bibfield  {journal}
  {\bibinfo  {journal} {J. Chem. Phys.}\ }\textbf {\bibinfo {volume} {151}},\
  \bibinfo {pages} {014113} (\bibinfo {year} {2019})}\BibitemShut {NoStop}%
\bibitem [{\citenamefont {Mátyus}\ and\ \citenamefont {Ferenc}()}]{MaFe22nad}%
  \BibitemOpen
  \bibfield  {author} {\bibinfo {author} {\bibfnamefont {E.}~\bibnamefont
  {Mátyus}}\ and\ \bibinfo {author} {\bibfnamefont {D.}~\bibnamefont
  {Ferenc}},\ }\href@noop {} {\bibfield  {journal} {\bibinfo  {journal} {Mol.
  Phys.}\ }\textbf {\bibinfo {volume} {120}}}\BibitemShut {NoStop}%
\bibitem [{\citenamefont {Staszewska}\ and\ \citenamefont
  {Wolniewicz}(2002)}]{StWo02}%
  \BibitemOpen
  \bibfield  {author} {\bibinfo {author} {\bibfnamefont {G.}~\bibnamefont
  {Staszewska}}\ and\ \bibinfo {author} {\bibfnamefont {L.}~\bibnamefont
  {Wolniewicz}},\ }\href {\doibase https://doi.org/10.1006/jmsp.2002.8546}
  {\bibfield  {journal} {\bibinfo  {journal} {J. Mol. Spectrosc.}\ }\textbf
  {\bibinfo {volume} {212}},\ \bibinfo {pages} {208} (\bibinfo {year}
  {2002})}\BibitemShut {NoStop}%
\bibitem [{\citenamefont {Wolniewicz}\ and\ \citenamefont
  {Staszewska}(2003)}]{WoSt03}%
  \BibitemOpen
  \bibfield  {author} {\bibinfo {author} {\bibfnamefont {L.}~\bibnamefont
  {Wolniewicz}}\ and\ \bibinfo {author} {\bibfnamefont {G.}~\bibnamefont
  {Staszewska}},\ }\href {\doibase
  https://doi.org/10.1016/S0022-2852(03)00121-8} {\bibfield  {journal}
  {\bibinfo  {journal} {J. Mol. Spectrosc.}\ }\textbf {\bibinfo {volume}
  {220}},\ \bibinfo {pages} {45} (\bibinfo {year} {2003})}\BibitemShut
  {NoStop}%
\bibitem [{\citenamefont {Siłkowski}\ \emph {et~al.}(2021)\citenamefont
  {Siłkowski}, \citenamefont {Zientkiewicz},\ and\ \citenamefont
  {Pachucki}}]{SiZsPa21}%
  \BibitemOpen
  \bibfield  {author} {\bibinfo {author} {\bibfnamefont {M.}~\bibnamefont
  {Siłkowski}}, \bibinfo {author} {\bibfnamefont {M.}~\bibnamefont
  {Zientkiewicz}}, \ and\ \bibinfo {author} {\bibfnamefont {K.}~\bibnamefont
  {Pachucki}},\ }in\ \href {\doibase
  https://doi.org/10.1016/bs.aiq.2021.05.012} {\emph {\bibinfo {booktitle} {New
  Electron Correlation Methods and their Applications, and Use of Atomic
  Orbitals with Exponential Asymptotes}}},\ \bibinfo {series} {Adv. Quant.
  Chem.}, Vol.~\bibinfo {volume} {83},\ \bibinfo {editor} {edited by\ \bibinfo
  {editor} {\bibfnamefont {M.}~\bibnamefont {Musial}}\ and\ \bibinfo {editor}
  {\bibfnamefont {P.~E.}\ \bibnamefont {Hoggan}}}\ (\bibinfo  {publisher}
  {Academic Press},\ \bibinfo {year} {2021})\ pp.\ \bibinfo {pages}
  {255--267}\BibitemShut {NoStop}%
\bibitem [{\citenamefont {Suzuki}\ and\ \citenamefont
  {Varga}(1998)}]{SuVaBook98}%
  \BibitemOpen
  \bibfield  {author} {\bibinfo {author} {\bibfnamefont {Y.}~\bibnamefont
  {Suzuki}}\ and\ \bibinfo {author} {\bibfnamefont {K.}~\bibnamefont {Varga}},\
  }\href@noop {} {\emph {\bibinfo {title} {Stochastic Variational Approach to
  Quantum-Mechanical Few-Body Problems}}}\ (\bibinfo  {publisher}
  {Springer-Verlag},\ \bibinfo {address} {Berlin},\ \bibinfo {year}
  {1998})\BibitemShut {NoStop}%
\bibitem [{\citenamefont {M\'atyus}\ and\ \citenamefont
  {Reiher}(2012)}]{MaRe12}%
  \BibitemOpen
  \bibfield  {author} {\bibinfo {author} {\bibfnamefont {E.}~\bibnamefont
  {M\'atyus}}\ and\ \bibinfo {author} {\bibfnamefont {M.}~\bibnamefont
  {Reiher}},\ }\href {\doibase 10.1063/1.4731696} {\bibfield  {journal}
  {\bibinfo  {journal} {J. Chem. Phys.}\ }\textbf {\bibinfo {volume} {137}},\
  \bibinfo {pages} {024104} (\bibinfo {year} {2012})}\BibitemShut {NoStop}%
\bibitem [{\citenamefont {Mátyus}(2019)}]{Ma19review}%
  \BibitemOpen
  \bibfield  {author} {\bibinfo {author} {\bibfnamefont {E.}~\bibnamefont
  {Mátyus}},\ }\href {\doibase 10.1080/00268976.2018.1530461} {\bibfield
  {journal} {\bibinfo  {journal} {Mol. Phys.}\ }\textbf {\bibinfo {volume}
  {117}},\ \bibinfo {pages} {590} (\bibinfo {year} {2019})}\BibitemShut
  {NoStop}%
\bibitem [{\citenamefont {Suzuki}\ \emph {et~al.}(1998)\citenamefont {Suzuki},
  \citenamefont {Usukura},\ and\ \citenamefont {Varga}}]{SuUsVa98}%
  \BibitemOpen
  \bibfield  {author} {\bibinfo {author} {\bibfnamefont {Y.}~\bibnamefont
  {Suzuki}}, \bibinfo {author} {\bibfnamefont {J.}~\bibnamefont {Usukura}}, \
  and\ \bibinfo {author} {\bibfnamefont {K.}~\bibnamefont {Varga}},\ }\href
  {\doibase 10.1088/0953-4075/31/1/007} {\bibfield  {journal} {\bibinfo
  {journal} {J. Phys. B: At. Mol. Opt. Phys.}\ }\textbf {\bibinfo {volume}
  {31}},\ \bibinfo {pages} {31} (\bibinfo {year} {1998})}\BibitemShut {NoStop}%
\bibitem [{Po0()}]{Po04}%
  \BibitemOpen
  \href@noop {} {}\bibinfo {note} {M. J. D. Powell, The NEWUOA software for
  unconstrained optimization without derivatives (DAMTP 2004/NA05), Report no.
  NA2004/08.}\BibitemShut {Stop}%
\bibitem [{\citenamefont {Kinghorn}\ and\ \citenamefont
  {Adamowicz}(1999)}]{KiAd99}%
  \BibitemOpen
  \bibfield  {author} {\bibinfo {author} {\bibfnamefont {D.~B.}\ \bibnamefont
  {Kinghorn}}\ and\ \bibinfo {author} {\bibfnamefont {L.}~\bibnamefont
  {Adamowicz}},\ }\href {\doibase 10.1103/PhysRevLett.83.2541} {\bibfield
  {journal} {\bibinfo  {journal} {Phys. Rev. Lett.}\ }\textbf {\bibinfo
  {volume} {83}},\ \bibinfo {pages} {2541} (\bibinfo {year}
  {1999})}\BibitemShut {NoStop}%
\bibitem [{\citenamefont {Kinghorn}\ and\ \citenamefont
  {Adamowicz}(2000)}]{KiAd00}%
  \BibitemOpen
  \bibfield  {author} {\bibinfo {author} {\bibfnamefont {D.~B.}\ \bibnamefont
  {Kinghorn}}\ and\ \bibinfo {author} {\bibfnamefont {L.}~\bibnamefont
  {Adamowicz}},\ }\href {\doibase 10.1063/1.1288376} {\bibfield  {journal}
  {\bibinfo  {journal} {J. Chem. Phys.}\ }\textbf {\bibinfo {volume} {113}},\
  \bibinfo {pages} {4203} (\bibinfo {year} {2000})}\BibitemShut {NoStop}%
\bibitem [{\citenamefont {Bubin}\ and\ \citenamefont
  {Adamowicz}(2003)}]{BuAd03}%
  \BibitemOpen
  \bibfield  {author} {\bibinfo {author} {\bibfnamefont {S.}~\bibnamefont
  {Bubin}}\ and\ \bibinfo {author} {\bibfnamefont {L.}~\bibnamefont
  {Adamowicz}},\ }\href {\doibase 10.1063/1.1537719} {\bibfield  {journal}
  {\bibinfo  {journal} {J. Chem. Phys.}\ }\textbf {\bibinfo {volume} {118}},\
  \bibinfo {pages} {3079} (\bibinfo {year} {2003})}\BibitemShut {NoStop}%
\bibitem [{\citenamefont {Sharkey}\ \emph
  {et~al.}(2013{\natexlab{a}})\citenamefont {Sharkey}, \citenamefont
  {Kirnosov},\ and\ \citenamefont {Adamowicz}}]{ShKiAd13}%
  \BibitemOpen
  \bibfield  {author} {\bibinfo {author} {\bibfnamefont {K.~L.}\ \bibnamefont
  {Sharkey}}, \bibinfo {author} {\bibfnamefont {N.}~\bibnamefont {Kirnosov}}, \
  and\ \bibinfo {author} {\bibfnamefont {L.}~\bibnamefont {Adamowicz}},\ }\href
  {\doibase 10.1103/PhysRevA.88.032513} {\bibfield  {journal} {\bibinfo
  {journal} {Phys. Rev. A}\ }\textbf {\bibinfo {volume} {88}},\ \bibinfo
  {pages} {032513} (\bibinfo {year} {2013}{\natexlab{a}})}\BibitemShut
  {NoStop}%
\bibitem [{\citenamefont {Sharkey}\ \emph
  {et~al.}(2013{\natexlab{b}})\citenamefont {Sharkey}, \citenamefont
  {Kirnosov},\ and\ \citenamefont {Adamowicz}}]{ShKiAd13b}%
  \BibitemOpen
  \bibfield  {author} {\bibinfo {author} {\bibfnamefont {K.~L.}\ \bibnamefont
  {Sharkey}}, \bibinfo {author} {\bibfnamefont {N.}~\bibnamefont {Kirnosov}}, \
  and\ \bibinfo {author} {\bibfnamefont {L.}~\bibnamefont {Adamowicz}},\ }\href
  {\doibase 10.1063/1.4826450} {\bibfield  {journal} {\bibinfo  {journal} {J.
  Chem. Phys.}\ }\textbf {\bibinfo {volume} {139}},\ \bibinfo {pages} {164119}
  (\bibinfo {year} {2013}{\natexlab{b}})}\BibitemShut {NoStop}%
\bibitem [{\citenamefont {Kirnosov}\ \emph {et~al.}(2015)\citenamefont
  {Kirnosov}, \citenamefont {Sharkey},\ and\ \citenamefont
  {Adamowicz}}]{KiShAd15}%
  \BibitemOpen
  \bibfield  {author} {\bibinfo {author} {\bibfnamefont {N.}~\bibnamefont
  {Kirnosov}}, \bibinfo {author} {\bibfnamefont {K.~L.}\ \bibnamefont
  {Sharkey}}, \ and\ \bibinfo {author} {\bibfnamefont {L.}~\bibnamefont
  {Adamowicz}},\ }\href {\doibase 10.1088/0953-4075/48/19/195101} {\bibfield
  {journal} {\bibinfo  {journal} {J. Phys. B: At. Mol. Opt. Phys.}\ }\textbf
  {\bibinfo {volume} {48}},\ \bibinfo {pages} {195101} (\bibinfo {year}
  {2015})}\BibitemShut {NoStop}%
\bibitem [{\citenamefont {Bubin}\ and\ \citenamefont
  {Adamowicz}(2006)}]{BuAd06}%
  \BibitemOpen
  \bibfield  {author} {\bibinfo {author} {\bibfnamefont {S.}~\bibnamefont
  {Bubin}}\ and\ \bibinfo {author} {\bibfnamefont {L.}~\bibnamefont
  {Adamowicz}},\ }\href {\doibase 10.1063/1.2204605} {\bibfield  {journal}
  {\bibinfo  {journal} {J. Chem. Phys.}\ }\textbf {\bibinfo {volume} {124}},\
  \bibinfo {pages} {224317} (\bibinfo {year} {2006})}\BibitemShut {NoStop}%
\bibitem [{\citenamefont {Bubin}\ and\ \citenamefont
  {Adamowicz}(2008)}]{BuAd08}%
  \BibitemOpen
  \bibfield  {author} {\bibinfo {author} {\bibfnamefont {S.}~\bibnamefont
  {Bubin}}\ and\ \bibinfo {author} {\bibfnamefont {L.}~\bibnamefont
  {Adamowicz}},\ }\href {\doibase 10.1063/1.2894866} {\bibfield  {journal}
  {\bibinfo  {journal} {J. Chem. Phys.}\ }\textbf {\bibinfo {volume} {128}},\
  \bibinfo {pages} {114107} (\bibinfo {year} {2008})}\BibitemShut {NoStop}%
\bibitem [{\citenamefont {Varga}(2019)}]{Va19}%
  \BibitemOpen
  \bibfield  {author} {\bibinfo {author} {\bibfnamefont {K.}~\bibnamefont
  {Varga}},\ }\href {\doibase 10.1103/PhysRevA.99.012504} {\bibfield  {journal}
  {\bibinfo  {journal} {Phys. Rev. A}\ }\textbf {\bibinfo {volume} {99}},\
  \bibinfo {pages} {012504} (\bibinfo {year} {2019})}\BibitemShut {NoStop}%
\bibitem [{\citenamefont {Muolo}\ \emph {et~al.}(2019)\citenamefont {Muolo},
  \citenamefont {Mátyus},\ and\ \citenamefont {Reiher}}]{MuMaRe19}%
  \BibitemOpen
  \bibfield  {author} {\bibinfo {author} {\bibfnamefont {A.}~\bibnamefont
  {Muolo}}, \bibinfo {author} {\bibfnamefont {E.}~\bibnamefont {Mátyus}}, \
  and\ \bibinfo {author} {\bibfnamefont {M.}~\bibnamefont {Reiher}},\ }\href
  {\doibase 10.1063/1.5121318} {\bibfield  {journal} {\bibinfo  {journal} {J.
  Chem. Phys.}\ }\textbf {\bibinfo {volume} {151}},\ \bibinfo {pages} {154110}
  (\bibinfo {year} {2019})}\BibitemShut {NoStop}%
\bibitem [{\citenamefont {Bubin}\ and\ \citenamefont
  {Adamowicz}(2020)}]{BuAd20}%
  \BibitemOpen
  \bibfield  {author} {\bibinfo {author} {\bibfnamefont {S.}~\bibnamefont
  {Bubin}}\ and\ \bibinfo {author} {\bibfnamefont {L.}~\bibnamefont
  {Adamowicz}},\ }\href {\doibase 10.1063/1.5144268} {\bibfield  {journal}
  {\bibinfo  {journal} {J. Chem. Phys.}\ }\textbf {\bibinfo {volume} {152}},\
  \bibinfo {pages} {204102} (\bibinfo {year} {2020})}\BibitemShut {NoStop}%
\bibitem [{\citenamefont {Bethe}\ and\ \citenamefont
  {Salpeter}(1975)}]{BeSabook75}%
  \BibitemOpen
  \bibfield  {author} {\bibinfo {author} {\bibfnamefont {H.~A.}\ \bibnamefont
  {Bethe}}\ and\ \bibinfo {author} {\bibfnamefont {E.~E.}\ \bibnamefont
  {Salpeter}},\ }\href@noop {} {\emph {\bibinfo {title} {Quantum Mechanics of
  One- and Two-Electron Systems}}}\ (\bibinfo  {publisher} {Springer},\
  \bibinfo {address} {Berlin, Germany},\ \bibinfo {year} {1975})\BibitemShut
  {NoStop}%
\bibitem [{\citenamefont {Pachucki}(2005)}]{Pa05FW}%
  \BibitemOpen
  \bibfield  {author} {\bibinfo {author} {\bibfnamefont {K.}~\bibnamefont
  {Pachucki}},\ }\href {\doibase 10.1103/PhysRevA.71.012503} {\bibfield
  {journal} {\bibinfo  {journal} {Phys. Rev. A}\ }\textbf {\bibinfo {volume}
  {71}},\ \bibinfo {pages} {012503} (\bibinfo {year} {2005})}\BibitemShut
  {NoStop}%
\bibitem [{\citenamefont {Araki}(1957)}]{Ar57}%
  \BibitemOpen
  \bibfield  {author} {\bibinfo {author} {\bibfnamefont {H.}~\bibnamefont
  {Araki}},\ }\href {\doibase 10.1143/PTP.17.619} {\bibfield  {journal}
  {\bibinfo  {journal} {Prog. of Theor. Phys.}\ }\textbf {\bibinfo {volume}
  {17}},\ \bibinfo {pages} {619} (\bibinfo {year} {1957})}\BibitemShut
  {NoStop}%
\bibitem [{\citenamefont {Sucher}(1958)}]{sucherPhD1958}%
  \BibitemOpen
  \bibfield  {author} {\bibinfo {author} {\bibfnamefont {J.}~\bibnamefont
  {Sucher}},\ }\href@noop {} {\enquote {\bibinfo {title} {Energy levels of the
  two-electron atom, to order $\alpha^3$ {Rydberg} ({PhD} dissertation,
  {C}olumbia {U}niversity)},}\ } (\bibinfo {year} {1958})\BibitemShut {NoStop}%
\bibitem [{\citenamefont {Komasa}\ \emph {et~al.}(2011)\citenamefont {Komasa},
  \citenamefont {Piszczatowski}, \citenamefont {Lach}, \citenamefont
  {Przybytek}, \citenamefont {Jeziorski},\ and\ \citenamefont
  {Pachucki}}]{KoPiLaPrJePa11}%
  \BibitemOpen
  \bibfield  {author} {\bibinfo {author} {\bibfnamefont {J.}~\bibnamefont
  {Komasa}}, \bibinfo {author} {\bibfnamefont {K.}~\bibnamefont
  {Piszczatowski}}, \bibinfo {author} {\bibfnamefont {G.}~\bibnamefont {Lach}},
  \bibinfo {author} {\bibfnamefont {M.}~\bibnamefont {Przybytek}}, \bibinfo
  {author} {\bibfnamefont {B.}~\bibnamefont {Jeziorski}}, \ and\ \bibinfo
  {author} {\bibfnamefont {K.}~\bibnamefont {Pachucki}},\ }\href {\doibase
  10.1021/ct900391p} {\bibfield  {journal} {\bibinfo  {journal} {J. Chem.
  Theory Comput.}\ }\textbf {\bibinfo {volume} {7}},\ \bibinfo {pages} {3105}
  (\bibinfo {year} {2011})}\BibitemShut {NoStop}%
\bibitem [{\citenamefont {Puchalski}\ \emph {et~al.}(2019)\citenamefont
  {Puchalski}, \citenamefont {Komasa}, \citenamefont {Czachorowski},\ and\
  \citenamefont {Pachucki}}]{PuKoCzPa19}%
  \BibitemOpen
  \bibfield  {author} {\bibinfo {author} {\bibfnamefont {M.}~\bibnamefont
  {Puchalski}}, \bibinfo {author} {\bibfnamefont {J.}~\bibnamefont {Komasa}},
  \bibinfo {author} {\bibfnamefont {P.}~\bibnamefont {Czachorowski}}, \ and\
  \bibinfo {author} {\bibfnamefont {K.}~\bibnamefont {Pachucki}},\ }\href
  {\doibase 10.1103/PhysRevLett.122.103003} {\bibfield  {journal} {\bibinfo
  {journal} {Phys. Rev. Lett.}\ }\textbf {\bibinfo {volume} {122}},\ \bibinfo
  {pages} {103003} (\bibinfo {year} {2019})}\BibitemShut {NoStop}%
\bibitem [{\citenamefont {Wang}\ and\ \citenamefont {Yan}(2018)}]{WaYa18}%
  \BibitemOpen
  \bibfield  {author} {\bibinfo {author} {\bibfnamefont {L.~M.}\ \bibnamefont
  {Wang}}\ and\ \bibinfo {author} {\bibfnamefont {Z.-C.}\ \bibnamefont {Yan}},\
  }\href {\doibase 10.1103/PhysRevA.97.060501} {\bibfield  {journal} {\bibinfo
  {journal} {Phys. Rev. A}\ }\textbf {\bibinfo {volume} {97}},\ \bibinfo
  {pages} {060501(R)} (\bibinfo {year} {2018})}\BibitemShut {NoStop}%
\bibitem [{\citenamefont {Ferenc}\ \emph {et~al.}(2020)\citenamefont {Ferenc},
  \citenamefont {Korobov},\ and\ \citenamefont {M\'atyus}}]{FeKoMa20}%
  \BibitemOpen
  \bibfield  {author} {\bibinfo {author} {\bibfnamefont {D.}~\bibnamefont
  {Ferenc}}, \bibinfo {author} {\bibfnamefont {V.~I.}\ \bibnamefont {Korobov}},
  \ and\ \bibinfo {author} {\bibfnamefont {E.}~\bibnamefont {M\'atyus}},\
  }\href {\doibase 10.1103/PhysRevLett.125.213001} {\bibfield  {journal}
  {\bibinfo  {journal} {Phys. Rev. Lett.}\ }\textbf {\bibinfo {volume} {125}},\
  \bibinfo {pages} {213001} (\bibinfo {year} {2020})}\BibitemShut {NoStop}%
\bibitem [{\citenamefont {Ireland}\ \emph {et~al.}(2021)\citenamefont
  {Ireland}, \citenamefont {Jeszenszki}, \citenamefont {M\'atyus},
  \citenamefont {Martinazzo}, \citenamefont {Ronto},\ and\ \citenamefont
  {Pollak}}]{IrJeMaMaRoPo21}%
  \BibitemOpen
  \bibfield  {author} {\bibinfo {author} {\bibfnamefont {R.}~\bibnamefont
  {Ireland}}, \bibinfo {author} {\bibfnamefont {P.}~\bibnamefont {Jeszenszki}},
  \bibinfo {author} {\bibfnamefont {E.}~\bibnamefont {M\'atyus}}, \bibinfo
  {author} {\bibfnamefont {R.}~\bibnamefont {Martinazzo}}, \bibinfo {author}
  {\bibfnamefont {M.}~\bibnamefont {Ronto}}, \ and\ \bibinfo {author}
  {\bibfnamefont {E.}~\bibnamefont {Pollak}},\ }\href {\doibase
  10.1021/acsphyschemau.1c00018} {\bibfield  {journal} {\bibinfo  {journal}
  {ACS Phys. Chem. Au}\ }\textbf {\bibinfo {volume} {2}},\ \bibinfo {pages}
  {23–37} (\bibinfo {year} {2021})}\BibitemShut {NoStop}%
\bibitem [{\citenamefont {Jeszenszki}\ \emph
  {et~al.}(2022{\natexlab{a}})\citenamefont {Jeszenszki}, \citenamefont
  {Ireland}, \citenamefont {Ferenc},\ and\ \citenamefont
  {Mátyus}}]{JeIrFeMa22}%
  \BibitemOpen
  \bibfield  {author} {\bibinfo {author} {\bibfnamefont {P.}~\bibnamefont
  {Jeszenszki}}, \bibinfo {author} {\bibfnamefont {R.~T.}\ \bibnamefont
  {Ireland}}, \bibinfo {author} {\bibfnamefont {D.}~\bibnamefont {Ferenc}}, \
  and\ \bibinfo {author} {\bibfnamefont {E.}~\bibnamefont {Mátyus}},\ }\href
  {\doibase https://doi.org/10.1002/qua.26819} {\bibfield  {journal} {\bibinfo
  {journal} {Int. J. Quant. Chem.}\ }\textbf {\bibinfo {volume} {122}},\
  \bibinfo {pages} {e26819} (\bibinfo {year} {2022}{\natexlab{a}})}\BibitemShut
  {NoStop}%
\bibitem [{\citenamefont {Jeszenszki}\ \emph {et~al.}(2021)\citenamefont
  {Jeszenszki}, \citenamefont {Ferenc},\ and\ \citenamefont
  {M\'atyus}}]{JeFeMa21}%
  \BibitemOpen
  \bibfield  {author} {\bibinfo {author} {\bibfnamefont {P.}~\bibnamefont
  {Jeszenszki}}, \bibinfo {author} {\bibfnamefont {D.}~\bibnamefont {Ferenc}},
  \ and\ \bibinfo {author} {\bibfnamefont {E.}~\bibnamefont {M\'atyus}},\
  }\href {\doibase 10.1063/5.0051237} {\bibfield  {journal} {\bibinfo
  {journal} {J. Chem. Phys.}\ }\textbf {\bibinfo {volume} {154}},\ \bibinfo
  {pages} {224110} (\bibinfo {year} {2021})}\BibitemShut {NoStop}%
\bibitem [{\citenamefont {Jeszenszki}\ \emph
  {et~al.}(2022{\natexlab{b}})\citenamefont {Jeszenszki}, \citenamefont
  {Ferenc},\ and\ \citenamefont {M\'atyus}}]{JeFeMa22}%
  \BibitemOpen
  \bibfield  {author} {\bibinfo {author} {\bibfnamefont {P.}~\bibnamefont
  {Jeszenszki}}, \bibinfo {author} {\bibfnamefont {D.}~\bibnamefont {Ferenc}},
  \ and\ \bibinfo {author} {\bibfnamefont {E.}~\bibnamefont {M\'atyus}},\
  }\href {\doibase 10.1063/5.0075096} {\bibfield  {journal} {\bibinfo
  {journal} {J. Chem. Phys.}\ }\textbf {\bibinfo {volume} {156}},\ \bibinfo
  {pages} {084111} (\bibinfo {year} {2022}{\natexlab{b}})}\BibitemShut
  {NoStop}%
\bibitem [{\citenamefont {Ferenc}\ \emph
  {et~al.}(2022{\natexlab{a}})\citenamefont {Ferenc}, \citenamefont
  {Jeszenszki},\ and\ \citenamefont {M\'atyus}}]{FeJeMa22}%
  \BibitemOpen
  \bibfield  {author} {\bibinfo {author} {\bibfnamefont {D.}~\bibnamefont
  {Ferenc}}, \bibinfo {author} {\bibfnamefont {P.}~\bibnamefont {Jeszenszki}},
  \ and\ \bibinfo {author} {\bibfnamefont {E.}~\bibnamefont {M\'atyus}},\
  }\href {\doibase 10.1063/5.0075097} {\bibfield  {journal} {\bibinfo
  {journal} {J. Chem. Phys.}\ }\textbf {\bibinfo {volume} {156}},\ \bibinfo
  {pages} {084110} (\bibinfo {year} {2022}{\natexlab{a}})}\BibitemShut
  {NoStop}%
\bibitem [{\citenamefont {Ferenc}\ \emph
  {et~al.}(2022{\natexlab{b}})\citenamefont {Ferenc}, \citenamefont
  {Jeszenszki},\ and\ \citenamefont {Matyus}}]{FeJeMa22b}%
  \BibitemOpen
  \bibfield  {author} {\bibinfo {author} {\bibfnamefont {D.}~\bibnamefont
  {Ferenc}}, \bibinfo {author} {\bibfnamefont {P.}~\bibnamefont {Jeszenszki}},
  \ and\ \bibinfo {author} {\bibfnamefont {E.}~\bibnamefont {Matyus}},\ }\href
  {\doibase 10.1063/5.0105355} {\bibfield  {journal} {\bibinfo  {journal} {J.
  Chem. Phys.}\ }\textbf {\bibinfo {volume} {157}},\ \bibinfo {pages} {094113}
  (\bibinfo {year} {2022}{\natexlab{b}})}\BibitemShut {NoStop}%
\bibitem [{\citenamefont {Wolniewicz}(1995)}]{Wo95Brel}%
  \BibitemOpen
  \bibfield  {author} {\bibinfo {author} {\bibfnamefont {L.}~\bibnamefont
  {Wolniewicz}},\ }\href {\doibase
  https://doi.org/10.1016/0009-2614(94)01493-F} {\bibfield  {journal} {\bibinfo
   {journal} {Chem. Phys. Lett.}\ }\textbf {\bibinfo {volume} {233}},\ \bibinfo
  {pages} {647} (\bibinfo {year} {1995})}\BibitemShut {NoStop}%
\bibitem [{\citenamefont {Wolniewicz}(1998)}]{Wo98EFrel}%
  \BibitemOpen
  \bibfield  {author} {\bibinfo {author} {\bibfnamefont {L.}~\bibnamefont
  {Wolniewicz}},\ }\href {\doibase 10.1063/1.476852} {\bibfield  {journal}
  {\bibinfo  {journal} {J. Chem. Phys.}\ }\textbf {\bibinfo {volume} {109}},\
  \bibinfo {pages} {2254} (\bibinfo {year} {1998})}\BibitemShut {NoStop}%
\bibitem [{\citenamefont {Pachucki}\ \emph {et~al.}(2005)\citenamefont
  {Pachucki}, \citenamefont {Cencek},\ and\ \citenamefont {Komasa}}]{PaCeKo05}%
  \BibitemOpen
  \bibfield  {author} {\bibinfo {author} {\bibfnamefont {K.}~\bibnamefont
  {Pachucki}}, \bibinfo {author} {\bibfnamefont {W.}~\bibnamefont {Cencek}}, \
  and\ \bibinfo {author} {\bibfnamefont {J.}~\bibnamefont {Komasa}},\ }\href
  {\doibase 10.1063/1.1888572} {\bibfield  {journal} {\bibinfo  {journal} {J.
  Chem. Phys.}\ }\textbf {\bibinfo {volume} {122}},\ \bibinfo {pages} {184101}
  (\bibinfo {year} {2005})}\BibitemShut {NoStop}%
\bibitem [{\citenamefont {Tiesinga}\ \emph {et~al.}(2021)\citenamefont
  {Tiesinga}, \citenamefont {Mohr}, \citenamefont {Newell},\ and\ \citenamefont
  {Taylor}}]{codata18}%
  \BibitemOpen
  \bibfield  {author} {\bibinfo {author} {\bibfnamefont {E.}~\bibnamefont
  {Tiesinga}}, \bibinfo {author} {\bibfnamefont {P.~J.}\ \bibnamefont {Mohr}},
  \bibinfo {author} {\bibfnamefont {D.~B.}\ \bibnamefont {Newell}}, \ and\
  \bibinfo {author} {\bibfnamefont {B.~N.}\ \bibnamefont {Taylor}},\ }\href
  {\doibase 10.1103/RevModPhys.93.025010} {\bibfield  {journal} {\bibinfo
  {journal} {Rev. Mod. Phys.}\ }\textbf {\bibinfo {volume} {93}},\ \bibinfo
  {pages} {025010} (\bibinfo {year} {2021})}\BibitemShut {NoStop}%
\bibitem [{\citenamefont {Wolniewicz}\ \emph {et~al.}(2006)\citenamefont
  {Wolniewicz}, \citenamefont {Orlikowski},\ and\ \citenamefont
  {Staszewska}}]{WoOrSt06}%
  \BibitemOpen
  \bibfield  {author} {\bibinfo {author} {\bibfnamefont {L.}~\bibnamefont
  {Wolniewicz}}, \bibinfo {author} {\bibfnamefont {T.}~\bibnamefont
  {Orlikowski}}, \ and\ \bibinfo {author} {\bibfnamefont {G.}~\bibnamefont
  {Staszewska}},\ }\href {\doibase https://doi.org/10.1016/j.jms.2006.04.020}
  {\bibfield  {journal} {\bibinfo  {journal} {J. Mol. Spectrosc.}\ }\textbf
  {\bibinfo {volume} {238}},\ \bibinfo {pages} {118} (\bibinfo {year}
  {2006})}\BibitemShut {NoStop}%
\bibitem [{\citenamefont {Wolniewicz}\ and\ \citenamefont
  {Dressler}(1992)}]{WoDr92}%
  \BibitemOpen
  \bibfield  {author} {\bibinfo {author} {\bibfnamefont {L.}~\bibnamefont
  {Wolniewicz}}\ and\ \bibinfo {author} {\bibfnamefont {K.}~\bibnamefont
  {Dressler}},\ }\href {\doibase 10.1063/1.462647} {\bibfield  {journal}
  {\bibinfo  {journal} {J. Chem. Phys.}\ }\textbf {\bibinfo {volume} {96}},\
  \bibinfo {pages} {6053} (\bibinfo {year} {1992})}\BibitemShut {NoStop}%
\bibitem [{\citenamefont {Dressler}\ and\ \citenamefont
  {Wolniewicz}(1995)}]{DrWo95}%
  \BibitemOpen
  \bibfield  {author} {\bibinfo {author} {\bibfnamefont {K.}~\bibnamefont
  {Dressler}}\ and\ \bibinfo {author} {\bibfnamefont {L.}~\bibnamefont
  {Wolniewicz}},\ }\href {\doibase https://doi.org/10.1002/bbpc.19950990303}
  {\bibfield  {journal} {\bibinfo  {journal} {Ber. Buns. Phys. Chem.}\ }\textbf
  {\bibinfo {volume} {99}},\ \bibinfo {pages} {246} (\bibinfo {year}
  {1995})}\BibitemShut {NoStop}%
\bibitem [{\citenamefont {Puchalski}\ \emph {et~al.}(2017)\citenamefont
  {Puchalski}, \citenamefont {Komasa},\ and\ \citenamefont
  {Pachucki}}]{PuKoPa17}%
  \BibitemOpen
  \bibfield  {author} {\bibinfo {author} {\bibfnamefont {M.}~\bibnamefont
  {Puchalski}}, \bibinfo {author} {\bibfnamefont {J.}~\bibnamefont {Komasa}}, \
  and\ \bibinfo {author} {\bibfnamefont {K.}~\bibnamefont {Pachucki}},\ }\href
  {\doibase 10.1103/PhysRevA.95.052506} {\bibfield  {journal} {\bibinfo
  {journal} {Phys. Rev. A}\ }\textbf {\bibinfo {volume} {95}},\ \bibinfo
  {pages} {052506} (\bibinfo {year} {2017})}\BibitemShut {NoStop}%
\bibitem [{\citenamefont {Yu}\ and\ \citenamefont {Dressler}(1994)}]{YuDr94}%
  \BibitemOpen
  \bibfield  {author} {\bibinfo {author} {\bibfnamefont {S.}~\bibnamefont
  {Yu}}\ and\ \bibinfo {author} {\bibfnamefont {K.}~\bibnamefont {Dressler}},\
  }\href {\doibase 10.1063/1.468263} {\bibfield  {journal} {\bibinfo  {journal}
  {J. Chem. Phys.}\ }\textbf {\bibinfo {volume} {101}},\ \bibinfo {pages}
  {7692} (\bibinfo {year} {1994})}\BibitemShut {NoStop}%
\bibitem [{\citenamefont {Hölsch}\ \emph {et~al.}(2018)\citenamefont
  {Hölsch}, \citenamefont {Beyer},\ and\ \citenamefont {Merkt}}]{HoBeMe18}%
  \BibitemOpen
  \bibfield  {author} {\bibinfo {author} {\bibfnamefont {N.}~\bibnamefont
  {Hölsch}}, \bibinfo {author} {\bibfnamefont {M.}~\bibnamefont {Beyer}}, \
  and\ \bibinfo {author} {\bibfnamefont {F.}~\bibnamefont {Merkt}},\ }\href
  {\doibase 10.1039/C8CP05233F} {\bibfield  {journal} {\bibinfo  {journal}
  {Phys. Chem. Chem. Phys.}\ }\textbf {\bibinfo {volume} {20}},\ \bibinfo
  {pages} {26837} (\bibinfo {year} {2018})}\BibitemShut {NoStop}%
\bibitem [{\citenamefont {Bai}\ \emph {et~al.}(2021)\citenamefont {Bai},
  \citenamefont {Zhong}, \citenamefont {Yan},\ and\ \citenamefont
  {Shi}}]{BaZhYaSh21}%
  \BibitemOpen
  \bibfield  {author} {\bibinfo {author} {\bibfnamefont {Z.-D.}\ \bibnamefont
  {Bai}}, \bibinfo {author} {\bibfnamefont {Z.-X.}\ \bibnamefont {Zhong}},
  \bibinfo {author} {\bibfnamefont {Z.-C.}\ \bibnamefont {Yan}}, \ and\
  \bibinfo {author} {\bibfnamefont {T.-Y.}\ \bibnamefont {Shi}},\ }\href
  {\doibase 10.1088/1674-1056/abc156} {\bibfield  {journal} {\bibinfo
  {journal} {Chin. Phys. B}\ }\textbf {\bibinfo {volume} {30}},\ \bibinfo
  {pages} {023101} (\bibinfo {year} {2021})}\BibitemShut {NoStop}%
\end{thebibliography}

\end{document}